\documentclass[12pt]{article}
\pdfoutput=1
\usepackage{amsmath,amssymb,latexsym,epsf,epsfig,graphicx,ifpdf}
\numberwithin{equation}{section}

\begin{document}

\newcommand{\sect}[1]{\setcounter{equation}{0}\section{#1}}
\renewcommand{\theequation}{\thesection.\arabic{equation}}
\newcommand{\prt}{\partial}
\newcommand{\II}{\mbox{${\mathbb I}$}}
\newcommand{\J}{\mbox{${\mathbb J}$}}
\newcommand{\CC}{\mbox{${\mathbb C}$}}
\newcommand{\RR}{\mbox{${\mathbb R}$}}
\newcommand{\QQ}{\mbox{${\mathbb Q}$}}
\newcommand{\ZZ}{\mbox{${\mathbb Z}$}}
\newcommand{\NN}{\mbox{${\mathbb N}$}}
\def\L{\mathbb L}
\def\UU{\mathbb U}
\def\S{\mathbb S}
\def\G{\mathbb G}
\def\tS{\widetilde{\mathbb S}}
\def\V{\mathbb V}
\def\tV{\widetilde{\mathbb V}}
\newcommand{\D}{{\mathbb D}}
\def\hint{H_{\rm int}}

\def\b1{\beta_1}

\newcommand{\rd}{{\rm d}}
\newcommand{\diag}{{\rm diag}}
\newcommand{\U}{{\cal U}}
\newcommand{\cH}{{\cal H}}
\newcommand{\cP}{{\cal P}}

\newcommand{\ph}{\varphi}
\newcommand{\phd}{\widetilde{\varphi}} 
\newcommand{\omt}{\widetilde{\Omega}} 
\newcommand{\phs}{\varphi^{(s)}}
\newcommand{\phb}{\varphi^{(b)}}
\newcommand{\phds}{\widetilde{\varphi}^{(s)}}
\newcommand{\phdb}{\widetilde{\varphi}^{(b)}}
\newcommand{\lambdad}{\widetilde{\lambda}}
\newcommand{\hC}{\widehat{C}}
\newcommand{\hL}{\widehat{L}}
\newcommand{\tx}{\widetilde{x}} 
\newcommand{\tA}{\widetilde{A}} 
\newcommand{\tB}{\widetilde{B}} 
\newcommand{\tO}{\widetilde{O}}
\newcommand{\phl}{\varphi_{i,L}}
\newcommand{\phr}{\varphi_{i,R}}
\newcommand{\phz}{\varphi_{i,Z}}
\newcommand{\mur}{\mu_{{}_R}}
\newcommand{\mul}{\mu_{{}_L}}
\newcommand{\muv}{\mu_{{}_V}}
\newcommand{\mua}{\mu_{{}_A}}

\def\a{\alpha}

\def\A{\mathcal A} 
\def\C{\mathcal C} 
\def\O{\mathcal O}
\def\E{\mathcal E} 
\def\T{\mathcal T} 
\def\I{\mathcal I}
\def\der{\partial }
\def\mis{{\frac{\rd k}{2\pi} }}
\def\ri{{\rm i}}
\def\xt{{\widetilde x}}
\def\ft{{\widetilde f}}
\def\gt{{\widetilde g}}
\def\qt{{\widetilde q}}
\def\tt{{\widetilde t}}
\def\tmu{{\widetilde \mu}}
\def\prt{{\partial}}
\def\tr{{\rm Tr}}
\def\inc{{\rm in}}
\def\out{{\rm out}}
\def\e{{\rm e}}
\def\eps{\varepsilon}
\def\k{\kappa}
\def\v{{\bf v}}

%%%%%%%%%%%%%%%%%%% INIZIO %%%%%%%%%%%%%%%%%%%%%%%

%%%%%%%% 
\newcommand{\finprf}{\null \hfill {\rule{5pt}{5pt}}\\[2.1ex]\indent}

%%%%%%%%%%%%%%%%%%%%%%%
%%%%%%%%%%%%%%%%%%%%%%%
\pagestyle{empty}
\rightline{October 2012}
%\rightline{Preliminary Draft}

\vfill

\begin{center}
{\Large\bf Luttinger Liquid in Non-equilibrium\\ Steady State}
\\[2.1em]

\bigskip

{\large Mihail Mintchev$^1$ and Paul Sorba$^2$}\\ 

\vskip 1.5 truecm 

{\it  
${}^1$ Istituto Nazionale di Fisica Nucleare and Dipartimento di Fisica dell'Universit\`a di Pisa, 
Largo Pontecorvo 3, 56127 Pisa, Italy\\ 
\bigskip 

${}^2$ Laboratoire de Physique Th\'eorique d'Annecy-le-Vieux, CNRS,\\  
9, Chemin de Bellevue, BP 110, F-74941 Annecy-le-Vieux Cedex, France}

\null

\noindent 

\vfill

\end{center}
\begin{abstract} 

We propose and investigate an exactly solvable model of non-equilibrium Luttinger 
liquid on a star graph, modeling a multi-terminal quantum wire junction. The boundary condition at the 
junction is fixed by an orthogonal matrix $\S$, which describes the splitting of the 
electric current among the leads. The system is driven away from equilibrium by 
connecting the leads to heat baths at different temperatures and chemical potentials. 
The associated non-equilibrium steady state depends on $\S$ and  
is explicitly constructed. In this context we develop a non-equilibrium bosonization 
procedure and compute some basic correlation functions.  
Luttinger liquids with general anyon statistics are considered. 
The relative momentum distribution away 
from equilibrium turns out to be the convolution of equilibrium anyon distributions 
at different temperatures. Both the charge and heat transport are 
studied. The exact current-current correlation function is derived and 
the zero-frequency noise power is determined.

\end{abstract}
\bigskip 
\medskip 
\bigskip 

\vfill
\rightline{LAPTH-047/12}
\rightline{IFUP-TH 20/2012}

\newpage
\pagestyle{plain}
\setcounter{page}{1}

\date{\today}

\section{Introduction}

The universal features of a large class of one-dimensional quantum models, exhibiting gapless 
excitations with linear spectrum, are successfully described \cite{Hald, Haldprl} by the 
Tomonaga-Luttinger (TL) liquid theory \cite{T50}-\cite{ML65}. This theory\footnote{For some more 
recent reviews we refer to \cite{voit-95}-\cite{egg-08}.} applies to 
various systems, including nanowire junctions and carbon nanotubes, which are available 
nowadays in experiment \cite{na1}-\cite{na3}. For this reason the study of 
non-equilibrium phenomena in the TL liquid phase attracts recently much 
attention \cite{GGM08}-\cite{Ines}. 

A typical non-equilibrium setup, considered in the literature, is the junction 
of two or more semi-infinite leads with electrons at different 
temperatures and/or chemical potentials. The junction is an interval of finite length $L$, where 
the electrons injected from the leads interact among themselves. This interaction drives 
the system away from equilibrium. Differently from the equilibrium TL liquid on the line, 
the non-equilibrium model defined in this way, is not exactly solvable. Nevertheless, it is 
extensively studied \cite{GGM08}-\cite{Ines} by various methods, 
including linear response theory, bosonization combined with the non-equilibrium 
Keldish formalism and perturbation theory. 

\begin{figure}[h]
\begin{center}
\begin{picture}(600,120)(40,250) 
\includegraphics[scale=0.8]{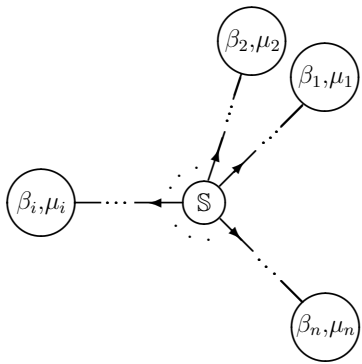}
\end{picture} 
\end{center}
\caption{A junction with scattering matrix $\S$ 
and $n$ semi-infinite leads, connected at infinity to thermal reservoirs with temperatures 
$\beta_i$ and chemical potentials $\mu_i$.} 
\label{fig1}
\end{figure}

One of the main goals of the present paper is to explore the possibility to construct and analyze 
an alternative {\it exactly solvable} model for a non-equilibrium TL junction. 
Since the universal features of such a system are expected to manifest themselves 
in the critical (scale invariant) limit, it is natural to shrink the domain of the non-equilibrium 
interaction to a point, taking $L\to 0$. For a complete description of the critical regime 
it is essential to take into account all point-like interactions, which ensure  
a unitary time evolution of the system. These interactions can be 
parametrized by a scattering matrix $\S$ localized in the junction point, as shown in the 
multi-terminal setup displayed in Fig.\ref{fig1}.  
Each lead contains a TL liquid, which at infinity is in contact with a heat reservoir with 
(inverse) temperature $\beta_i$ and chemical potential $\mu_i$. Our first step below is to show
that there exists a non-equilibrium steady state (NESS), which describes the TL configuration 
in Fig.\ref{fig1}. This state is characterized by non-trivial time independent electric 
and heat currents, flowing in the leads. The scattering matrix $\S$ is implemented 
by imposing specific boundary conditions at the junction. It turns out that the boundary 
conditions, which describe the splitting of the electric steady current in the junction, 
lead to an exactly solvable problem. In fact, we establish the operator solution in this case 
and investigate the relative non-equilibrium correlation functions in the NESS representation. 

The TL theory has been introduced originally 
\cite{T50}-\cite{ML65} for describing fermion systems. 
It has been understood later on \cite{Liguori:1999tw}-\cite{Calabrese:2007ty} that 
the fermion TL liquid is actually an element of a more general family of {\it anyon} TL 
liquids\footnote{Similar results hold \cite{PKA09, Calabrese:2008fv} for the 
Lieb-Lineger and Calogero-Sutherland models.}, which obey Abelian 
braid statistics. In this paper we explore the general anyon TL liquid, 
obtaining the conventional fermionic and bosonic ones as a special cases. 

{}From the two-point anyon correlation functions we extract the 
NESS distribution of the TL anyon excitations. In 
momentum space this non-equilibrium distribution is a {\it nested convolution} of 
equilibrium distributions at different temperatures and chemical potentials. As expected, 
the convolution depends on the scattering matrix $\S$, which drives the system away from 
equilibrium. We investigate also the NESS correlators of the electric and energy currents, 
describing in detail the charge and heat transport in the junction. The zero-frequency 
noise power is deduced from the two-point current-current correlation function, whose 
exact expression in terms of hypergeometric functions is established. We prove the breakdown of 
time reversal invariance as well. 

The paper has the following structure. In the next section we construct non-equilibrium 
chiral fields in a NESS on a star graph modeling the junction. 
We derive here the non-equilibrium Casimir energy and the heat current and 
compare the latter with the conformal field theory result. 
In section 3 we develop a non-equilibrium finite temperature operator 
bosonization procedure. We also establish the operator solution, subject to the current 
splitting boundary condition at the junction. We show that this condition covers 
two different physical situations, corresponding to a junction with and without charge dissipation.  
The non-equilibrium correlation functions are investigated in section 4, where 
the anyon NESS distributions are derived. The charge and heat transport as well as the 
noise are also studied there. Section 5 provides a concise outlook of the paper and contains some general 
observations. The appendix collects some results about the asymptotic properties of 
the anyon NESS correlators.

\section{Non-equilibrium chiral fields on a star graph} 

\begin{figure}[h]
\begin{center}
\begin{picture}(500,70)(-150,20) 
\includegraphics[scale=1]{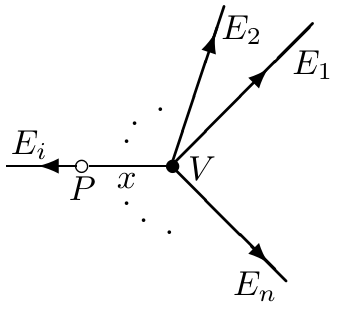}
\end{picture} 
\end{center}
\caption{A star graph $\Gamma$ with $n$ edges modelling the junction of $n$ quantum wires.}
\label{fig2}
\end{figure}

The fundamental building blocks of bosonization away from equilibrium are the free 
massless scalar field $\ph $ and its dual $\phd $. The fields $\ph $ and $\phd $ 
propagate on a star graph $\Gamma$, which is shown in Fig. \ref{fig2} and models the quantum wire junction. 
The edges $E_i$ are half-lines and each point $P$ in the bulk $\Gamma \setminus V$ 
of $\Gamma$ is uniquely determined by its coordinates $(x,i)$, where $x > 0$ is the 
distance to the vertex $V$ and $i=1,...,n$ labels the edge. Besides the massless Klein-Gordon 
equation, the fields $\ph $ and $\phd $ satisfy the duality relations
\begin{equation}
\der_t \phd (t,x,i) = - \der_x \ph (t,x,i) \, ,\qquad 
\der_x \phd (t,x,i) = - \der_t \ph (t,x,i) \, .
\label{dual}
\end{equation} 
The initial conditions are fixed by the equal-time canonical commutation relations 
\begin{equation}
[\ph (t,x,i)\, ,\, \ph (t,y,j)]_{{}_-} =
[\phd (t,x,i)\, ,\, \phd (t,y,j)]_{{}_-} = 0 \, ,
\label{ecc1}
\end{equation}
\begin{equation}
[(\der_t\ph ) (t,x,i)\, ,\, \ph (t,y,j)]_{{}_-}  =
[(\der_t\phd )(t,x,i)\, ,\, \phd (t,y,j)]_{{}_-}  =
-\ri \delta_{ij}\delta (x-y) \, .
\label{ecc2}
\end{equation} 
In order to determine the dynamics completely, one must impose some boundary conditions 
at the vertex $x=0$. These conditions are conveniently formulated in terms of the combinations 
\begin{equation}
\phr (t-x) = \ph (t,x,i) + \phd (t,x,i)\, , \qquad 
\phl (t+x) = \ph (t,x,i) - \phd (t,x,i)\, ,
\label{chi1}
\end{equation}
which depend on $t-x$ and $t+x$ respectively and define right and left chiral fields $\phz$ on $\Gamma$. 
The most general {\it scale invariant} boundary conditions, generating a 
{\it unitary time evolution} of $\ph$ and $\phd$, are parametrized by the orthogonal group $O(n)$ 
and read \cite{ks-00}-\cite{Bellazzini:2008mn} 
\begin{equation}
\phr (\xi ) = \sum_{j=1}^n \S_{ij}\, \ph_{j,L}(\xi ) \, , \qquad \S \in O(n)\, .  
\label{bc1}
\end{equation}
These simple conditions capture the universal features of the system and 
$\S$ has a straightforward physical interpretation: 
the vertex $V$ of $\Gamma$ represents a {\it scale invariant point-like defect}, 
$\S$ being the associated scattering matrix.

\subsection{The non-equilibrium steady state $\Omega_{\beta,\mu_{{}_b}}$}

Our next step is to construct a steady state $\Omega_{\beta,\mu_{{}_b}}$, which 
captures the evolution of the chiral fields $\phz$ on $\Gamma$, whose edges 
are attached at infinity to thermal reservoirs at inverse temperatures $\beta_i$ as shown 
in Fig. \ref{fig1}. In the boson case we take all chemical potentials to be 
equal\footnote{This choice will not prevent us to deal in the fermion case below 
with arbitrary $\mu_i$.}, setting $\mu_i = \mu_b$ in all reservoirs. The system is 
away from equilibrium if $\S$ contains at least one non-trivial transmission coefficient among 
reservoirs with different temperature. The construction of $\Omega_{\beta,\mu_{{}_b}}$, 
described below, follows the scheme developed in \cite{Mintchev:2011mx} and is 
based on scattering theory. It adapts to the case under consideration some 
modern ideas \cite{els-96}-\cite{st-06} about NESS.  
The framework is purely algebraic and generalizes the definition \cite{BR} of equilibrium Gibbs 
state over the algebra of canonical commutation relations (CCR). 

We start by observing that the massless Klein-Gordon equation 
and the relations (\ref{dual}) lead to the following representation 
\begin{eqnarray}
&&\phr (\xi) = \int_0^\infty  \frac{\rd k}{\pi \sqrt{2}}\, \sqrt {\Delta_\lambda (k)} 
\left [a_i^\ast (k) \e^{\ri k \xi} + a_i(k)\e^{-\ri k \xi}\right ] \, , 
\label{fil1} \\
&& \phl (\xi) = \int_0^\infty  \frac{\rd k}{\pi \sqrt{2}}\, \sqrt {\Delta_\lambda (k)}
\left [a_i^\ast (-k) \e^{\ri k \xi} + a_i(-k) \e^{-\ri k \xi}\right ] \, , 
\label{fil2}
\end{eqnarray}
$\Delta_\lambda $ being some distribution to be fixed below. Using that $\S$ is a real matrix, the boundary 
condition (\ref{bc1}) implies the constraints 
\begin{equation} 
a_i(k) = \sum_{j=1}^n S_{ij}(k)\, a_j (-k) \, , \qquad 
a^\ast_i (k) = \sum_{j=1}^n  S_{ij}(k)\, a^\ast_ j(-k)\, ,    
\label{constr1}
\end{equation} 
where 
\begin{equation}
S(k) = \theta(-k) \S^t + \theta (k) \S\, ,  
\label{S}
\end{equation} 
$\theta$ is the Heaviside step function and $\S^t$ indicates the transpose of $\S$. 
{}From the equal-time commutation relations (\ref{ecc1}, \ref{ecc2}) one infers that the elements 
$\{a_i(k),\, a^*_i(k)\, :\, k \in \RR, \, i=1,...,n\}$ generate the following deformation $\A$ of the 
algebra of CCR: 
\begin{equation} 
[a_i(k)\, ,\, a_j(p)]_{{}_-} = [a^*_i (k)\, ,\, a^*_j (p)]_{{}_-} = 0 \, , 
\label{rta1}
\end{equation}
\begin{equation}
[a_i(k)\, ,\, a^*_j (p)]_{{}_-} = 2\pi [\delta (k-p)\delta_{ij} + S_{ij}(k)\delta(k+p)] \, . 
\label{rta2}
\end{equation}  
Moreover, (\ref{ecc1}, \ref{ecc2}) imply that 
\begin{equation}
|k| \Delta_\lambda (k) = 1\, . 
\label{distr1}
\end{equation}
There exist a one-parameter family of tempered 
distributions, which solve this equation in $\RR$. A convenient representation of this family is 
given by \cite{Liguori:1997vd}
\begin{equation}
\Delta_\lambda (k) = \frac{\rd }{\rd k} \left [\theta (k) \ln \frac {k}{\lambda}\right ]\, ,
\label{distr2}
\end{equation}
where $\lambda >0$ is a free parameter with dimension of mass, having well-known infrared origin. 

The above structure is very general and equations (\ref{fil1},\ref{fil2}) apply to any 
representation of the algebra $\A$, which is a simplified version of the so called 
reflection-transmission (RT) algebra \cite{Liguori:1996xr}-\cite{Mintchev:2003ue}, 
describing factorized scattering in integrable models with point-like defects in one dimension. 
The Fock and the Gibbs state over $\A$ describe equilibrium physics 
and have been largely explored.  We will investigate here the NESS 
$\Omega_{\beta,\mu_{{}_b}}$, which describes the physical situation shown in Fig. \ref{fig1}. For 
this purpose we first observe that the sub-algebras $\A_\inc$ and $\A_\out$, 
generated by the elements $\{a_i(k),\, a^*_i(k)\, :\, k<0\}$ 
and $\{a_i(k),\, a^*_i(k)\, :\, k>0\}$ respectively, parametrize the {\it asymptotic incoming} and 
{\it outgoing} fields. 
Accordingly, both $\A_\inc$ and $\A_\out$ are conventional CCR algebras; in fact  
the $\delta (k+p)$ term in (\ref{rta2}) vanishes if both momenta are negative or positive. 
It is worth stressing that (\ref{constr1}) relate $\A_\inc$ with $\A_\out$ and that the 
whole RT algebra $\A$ can be generated via (\ref{constr1}) either by $\A_\inc$, or by $\A_\out$. 
The main idea for constructing $\Omega_{\beta , \mu_b}$ is based on this kind of 
asymptotic completeness property. Starting with an equilibrium state 
on $\A_\inc$, we will extend it by means of (\ref{constr1}) to a non-equilibrium state on 
the whole algebra $\A$. For this purpose we introduce the edge Hamiltonian and number operators 
\begin{equation} 
h_i = \int_{-\infty}^0 \frac{\rd k}{2\pi} |k| a^*_i(k) a_i(k)\, , \qquad 
n_i = \int_{-\infty}^0 \frac{\rd k}{2\pi} a^*_i(k) a_i(k)\, ,   
\label{ac}
\end{equation} 
which describe the {\it asymptotic} dynamics at $t=-\infty$ (i.e. before the interaction) 
in terms of $\A_\inc$. Defining   
\begin{equation} 
K = \sum_{i=1}^n \beta_i (h_i -\mu_b n_i) \, , \qquad \beta_i \geq 0\, , 
\label{kop}
\end{equation} 
we introduce the equilibrium Gibbs state over $\A_\inc$ in the standard way \cite{BR}. For 
any polynomial $\cP$ over $\A_\inc$ we set 
\begin{eqnarray}
\left (\Omega_{\beta , \mu_b}\, ,\,  \cP\bigl (a_i^*(k_i), a_j(p_j)\bigr ) \Omega_{\beta , \mu_b} \right ) \equiv 
\langle \cP\bigl (a_i^*(k_i), a_j(p_j))\bigr ) \rangle_{\beta, \mu_b} = 
\nonumber \\
\frac{1}{Z} \tr \left [\e^{-K} \cP\bigl (a_i^*(k_i), a_j(p_j)\bigr )\right ]\, ,  
\qquad \qquad \qquad 
\label{def1}
\end{eqnarray} 
where $k_i<0,\; p_j<0$ and $Z = \tr \left ( \e^{-K}\right )$. 
All the expectation values (\ref{def1}) can be computed \cite{BR} 
by purely algebraic manipulations and can be expressed in 
terms of the two-point functions, which are written in terms of 
the familiar Bose distribution 
\begin{equation} 
b_i (k) = \frac{\e^{-\beta_i [|k| - \mu_b]}}{1- \e^{-\beta_i [|k| - \mu_b]}} 
\label{fbd1} 
\end{equation}
in the following way 
\begin{equation}
\langle a_j^*(p)a_i(k) \rangle_{\beta, \mu_b} = 
b_i(k) \delta_{ij} 2\pi \delta (k-p)\, ,  
\label{2a}
\end{equation}
\begin{equation} 
\langle a_i(k)a_j^*(p)\rangle_{\beta, \mu_b} = [1+b_i(k)] 
\delta_{ij} 2\pi \delta (k-p)\, .  
\label{2b}
\end{equation}
We stress that (\ref{2a},\ref{2b}) hold on $\A_\inc$, i.e. only for {\it negative} momenta. 
The common for all reservoirs chemical potential $\mu_b<0$ allows to avoid 
in (\ref{2a},\ref{2b}) the infrared singularity at $k=0$. 
We anticipate that $\mu_b$ has nothing to do with the {\it fermion} chemical 
potentials, appearing in the non-equilibrium bosonization procedure described in the next section, 
where the limit $\mu_b \to 0^-$ exist and will be performed.  
 
The next step is to extend (\ref{def1}-\ref{2b}) to 
the whole RT algebra $\A$, namely to {\it positive} momenta. Employing (\ref{constr1}) one finds  
\begin{eqnarray} 
\langle a_j^*(p)a_i(k)\rangle_{\beta, \mu_b} = 
2\pi \Bigl\{\Bigl [\theta(-k)b_i(k) \delta_{ij}+ 
\theta(k)\sum_{l=1}^n \S_{il}\, b_l(k)\, \S^t_{lj}\Bigr ] \delta (k-p)  
\nonumber \\
+ \Bigl [\theta(-k)b_i(k)\, \S^t_{ij} + \theta(k)\S_{ij}\, b_j(k) \Bigr ] \delta (k+p) \Bigr\}\, .
\qquad \; \; \,  
\label{cor1}
\end{eqnarray} 
The expression for $\langle a_i(k)a_j^*(p)\rangle_{\beta, \mu}$ is 
obtained from (\ref{cor1}) by the substitution 
\begin{equation} 
b_i(k) \longmapsto 1+b_i(k) =  \frac{1}{1- \e^{-\beta_i [|k| - \mu_b]}} \, . 
\label{fbd2} 
\end{equation} 

The final step is to compute a generic correlation function. By means of the commutation 
relations (\ref{rta1},\ref{rta2}), this problem is reduced to the evaluation of correlators of the form 
$\langle \prod_{m=1}^M a_{i_m}(k_{i_m}) \prod_{n=1}^N a^\ast_{j_n}(p_{j_n})\rangle_{\beta,\mu_b}$, 
which can be computed by iteration via 
\begin{eqnarray} 
\langle \prod_{m=1}^M a_{i_m}(k_{i_m}) \prod_{n=1}^N a^{\ast j_n}(p_{j_n})\rangle_{\beta,\mu_b} = 
\qquad \qquad \qquad \nonumber \\
\delta_{MN}\, \sum_{m=1}^M \langle a_{i_1}(k_{i_1})a^{\ast j_m}(p_{j_m})\rangle_{\beta,\mu_b}  
\, \langle \prod_{m=2}^M a_{i_m}(k_{i_m}) 
\prod_{n=1\atop {n\not=m} }^N a^{\ast j_n}(p_{j_n}) \rangle_{\beta,\mu_b} \, .   
\label{gcf2}
\end{eqnarray} 

We would like to mention in conclusion that the use of the RT algebra 
$\A$ in the construction of $\Omega_{\beta, \mu_b}$ 
represents only a convenient choice of coordinates, which has a simple physical 
interpretation in terms of scattering data and applies 
to a variety of systems \cite{Mintchev:2011mx, Caudrelier:2012xy} with point-like defects. 
\bigskip 

\subsection{Energy density and energy transport in $\Omega_{\beta,\mu_b}$} 

In order to illustrate the physical properties of $\Omega_{\beta,\mu_b}$, 
it is instructive to investigate the non-equilibrium energy density and transport 
associated with the scalar field $\ph$. The equations of motion imply the conservation  
\begin{equation} 
\der_t \theta_{tt}(t,x,i)  - \der_x \theta_{xt}(t,x,i) = 0\, , 
\label{econs}
\end{equation}
of the energy-momentum tensor 
\begin{eqnarray} 
\theta_{tt}(t,x,i) &=& \frac{1}{2}: \left [ (\prt_t \ph )(\prt_t \ph ) - 
\ph (\prt_x^2 \ph ) \right ]:(t,x,i)\, , 
\label{emtph1}\\
\theta_{xt}(t,x,i) &=& \frac{1}{2} :\left [ (\prt_x \ph)( \prt_t \ph) - 
\ph (\prt_x \prt_t \ph )\right ] :(t,x,i)\, , 
\label{emtph2}
\end{eqnarray} 
where $:\cdots :$ denotes the normal product in the algebra $\A$. The boundary condition 
(\ref{bc1}) implies the Kirchhoff rule 
\begin{equation} 
\sum_{i=1}^n \theta_{xt} (t,0,i) = 0\, , 
\label{K1}
\end{equation}
which, combined with (\ref{econs}), ensures energy conservation. 

The derivation of $\langle \theta_{tt}(t,x,i)\rangle_{\beta,\mu_b}$ and 
$\langle \theta_{xt}(t,x,i)\rangle_{\beta,\mu_b}$ is based on the expectation value 
\begin{eqnarray} 
\langle : \ph(t_1,x_1,i_1) \ph(t_2,x_2,i_2): \rangle_{\beta,\mu_b} = \qquad \qquad \qquad 
\nonumber \\
\int_0^\infty \frac{\rd k}{2\pi}\Delta_\lambda (k) \Bigl \{\delta_{i_1i_2} b_{i_1}(k) \cos[k(t_{12}+x_{12}] + 
\S_{i_1i_2} b_{i_2}(k) \cos[k(t_{12}-\tx_{12}] + 
\nonumber \\
b_{i_1}(k)\, \S^t_{i_1i_2} \cos[k(t_{12}+\tx_{12}] 
+\sum_{j=1}^n \S_{i_1j}\, b_j(k)\, \S^t_{ji_2}  \cos[k(t_{12}-x_{12}]\Bigr \}\, , 
\label{ncf}
\end{eqnarray} 
where $\tx_{12} = x_1+x_2$. 
Plugging (\ref{ncf}) in the definitions (\ref{emtph1},\ref{emtph2}), one gets 
\begin{eqnarray} 
\E_i(\beta, \mu_b) \equiv 
\langle \theta_{tt}(t,x,i)\rangle_{\beta,\mu_b} = \qquad \qquad \qquad \qquad 
\nonumber \\
\S_{ii} \int_0^\infty  \frac{\rd k}{\pi}\, k\, \cos(2kx) b_i(k)+ \sum_{j=1}^n \left (\delta_{ij}+\S_{ij}^2 \right )
\frac{1}{2\pi \beta_j^2}\, {\rm Li}_2 \left (\e^{\beta_j \mu_b}\right )\, , \; \; \,   
\label{tr1}\\
\T_i(\beta, \mu_b) \equiv \langle \theta_{xt}(t,x,i)\rangle_{\beta,\mu_b} = 
\sum_{j=1}^n \left (\delta_{ij}-\S_{ij}^2\right ) 
\frac{1}{2\pi \beta_j^2}\, {\rm Li}_2 \left (\e^{\beta_j \mu_b}\right )\, , \; \; \, 
\label{tr2}
\end{eqnarray}
where ${\rm Li}_2$ is the dilogarithm function. Eq. (\ref{tr1}) determines the Casimir energy, 
whereas (\ref{tr2}) describes the energy (heat) transport. Both are time-independent, 
thus confirming that we are dealing with a steady state. Notice also that the 
energy density is $x$-dependent, which reflects the breaking of 
translation invariance by the junction and is consistent with the conservation law (\ref{econs}). 
Since $\S$ is an orthogonal matrix, $\T_i(\beta,\mu_b)$ obviously satisfies the Kirchhoff's rule (\ref{K1}). 
The energy density $\E_i(\beta, \mu_b)$ can be written in the equivalent form 
\begin{equation} 
\E_i(\beta, \mu_b) = 
\S_{ii} \int_0^\infty  \frac{\rd k}{\pi}\, k\, \cos(2kx) b_i(k) + 
\frac{1}{\pi \beta_i^2}\, {\rm Li}_2 \left (\e^{\beta_i \mu_b}\right )
- \sum_{j=1}^n \left (\delta_{ij}-\S_{ij}^2 \right )
\frac{1}{2\pi \beta_j^2}\, {\rm Li}_2 \left (\e^{\beta_j \mu_b}\right )\, ,
\label{ntr}
\end{equation}
where the last term vanishes at equilibrium and describes therefore the {\it non-equilibrium 
contribution} to the Casimir energy. 

{}For a junction with $n=2$ wires there are two 
one-parameter families 
\begin{equation} 
\S^+ = \left(\begin{array}{cc}\cos \theta & \sin \theta \\ 
-\sin \theta  & \cos \theta  \\ \end{array} \right)\, ,   \qquad 
\S^- = \left(\begin{array}{cc}\cos \theta  & \sin \theta \\ 
\sin \theta & -\cos \theta \\ \end{array} \right)\, , \quad \theta \in [0,2\pi)\, , 
\label{ex1}
\end{equation}
of scattering matrices with ${\rm det} (\S^\pm) = \pm 1$. For both families one finds  
\begin{equation}
\T_1(\beta, \mu_b) = - \T_2(\beta, \mu_b) = 
\frac{\sin^2 \theta }{2\pi} \left [ \frac{1}{\beta_1^2} {\rm Li}_2 \left (\e^{\beta_1 \mu_b}\right ) - 
\frac{1}{\beta_2^2} {\rm Li}_2 \left (\e^{\beta_2 \mu_b}\right )\right ]\, , 
\label{ctr1}
\end{equation}
which deserves a comment. The heath transport has been investigated recently 
in the framework of conformal field theory (CFT) for generic central charge $c$ in \cite{bd-12}. 
Eq. (\ref{ctr1}) confirms the result of \cite{bd-12} for $c=1$ and extends this result in two directions:
imperfect junction with transmission probability $\sin^2(\theta)$ and $\mu_b\not=0$. 
We observe in this respect that derivation of the dilogarithm terms in (\ref{ctr1}) is problematic 
in a CFT context, because both $\beta_1 \mu_b$ and $\beta_2 \mu_b$ are 
non-trivial dimensionless parameters.     

In the limit $\mu_b \to 0^-$ the $x$-dependent integral in (\ref{tr1}) can be evaluated explicitly 
and one finds 
\begin{eqnarray} 
\E_i(\beta, 0) &=&\frac{\pi}{12}\sum_{j=1}^n \left (\delta_{ij}+\S^2_{ij} \right )\frac{1}{\beta_j^2} + 
\S_{ii}\frac{1}{8\pi x^2} - \S_{ii}\frac{\pi}{2\beta_i^2 \left [\sinh \left (2\pi \frac{x}{\beta_i}\right )\right ]^2}\, , 
\label{tr3} \\
\T_i(\beta, 0) &=& \frac{\pi}{12}\sum_{j=1}^n \left (\delta_{ij} - \S_{ij}^2 \right )
\frac{1}{\beta_j^2}\, . \qquad 
\label{tr4}
\end{eqnarray}

\subsection{Chiral NESS correlators} 

In the non-equilibrium bosonization procedure, developed below, we will need the 
correlation functions of the chiral fields (\ref{fil1},\ref{fil2})) in the NESS $\Omega_{\beta, \mu_b}$. 
It is easily seen that all of them can be expressed in terms of the distribution 
\begin{equation} 
w (\xi,\beta;\lambda, \mu_b)= 
\int_0^\infty \frac{\rd k}{\pi} \Delta_\lambda (k) \left [\frac{\e^{-\beta [k - \mu_b]}}{1- \e^{-\beta [k - \mu_b]}} 
\e^{\ri k \xi} + \frac{1}{1- \e^{-\beta [k - \mu_b]}} 
\e^{-\ri k \xi}\right ]\, . 
\label{w}
\end{equation} 
The full $\lambda$-dependence and the singularity at $\mu_b=0$ of (\ref{w}) are captured by \cite{Liguori:1999tw}  
\begin{equation} 
w (\xi,\beta;\lambda, \mu_b) = \frac{1}{\pi} 
\left \{ \frac{2}{\beta |\mu_b|} \ln \frac{|\mu_b|}{\lambda} - 
\ln \left [2 \ri\, \sinh \left (\frac{\pi}{\beta} \xi - \ri \varepsilon \right ) \right ]\right \} + o(\mu_b)\, , 
\label{wa}
\end{equation} 
where $o(\mu_b)$ stands for $\lambda$-independent terms, which 
vanish in the limit $\mu_b \to 0^-$. It is convenient at this point to relate 
the (up to now free) infrared regularization parameter $\lambda $ to $\mu_b$ by means of  
\begin{equation}
\lambda = |\mu_b| \, .
\label{lambdamu}
\end{equation} 
The limit $\mu_b \to 0^-$ in (\ref{w}) now exists and gives the 
distribution\footnote{The $\ri \varepsilon$ prescription, adopted 
throughout the paper, indicates as usual the weak limit $\varepsilon \to 0^+$.} 
\begin{equation} 
w (\xi,\beta )\equiv \lim_{\mu_b \to 0^-} w (\xi,\beta; \lambda = |\mu_b|, \mu_b) = 
-\frac{1}{\pi} \ln \left [2 \ri\, \sinh \left (\frac{\pi}{\beta} \xi - \ri \varepsilon \right ) \right ] \, ,  
\label{w1}
\end{equation} 
which is the fundamental block of the NESS chiral correlation functions. In fact, for the 
two-point correlators one gets 
\begin{eqnarray}
\langle \ph_{i_1,L}(\xi_1) \ph_{i_2,L}(\xi_2) \rangle_\beta &=& 
\delta_{i_1i_2}\, w (\xi_{12},\beta_{i_1})\, , 
\label{ll}\\
\langle \ph_{i_1,L}(\xi_1) \ph_{i_2,R}(\xi_2) \rangle_{\beta} &=& 
w (\xi_{12},\beta_{i_1})\, \S^t_{i_1i_2}\, , 
\label{lr} \\
\langle \ph_{i_1,R}(\xi_1) \ph_{i_2,L}(\xi_2) \rangle_{\beta} &=& 
\S_{i_1i_2}\, w (\xi_{12},\beta_{i_2})\, , 
\label{rl} \\ 
\langle \ph_{i_1,R}(\xi_1) \ph_{i_2,R}(\xi_2) \rangle_{\beta} &=& 
\sum_{j=1}^n \S_{i_1j}\, w (\xi_{12},\beta_j)\, \S^t_{ji_2}\, . 
\label{rr}
\end{eqnarray} 
where $\xi_{12}\equiv \xi_1-\xi_2$. As expected, the point-like interaction 
in the vertex of $\Gamma$ induce a non-trivial left-right mixing described by (\ref{lr},\ref{rl}).

\section{Bosonization away from equilibrium} 

The possibility to express fermions in terms of bosons in 1+1 dimensional space-time 
has been discovered long ago by Jordan and Wigner \cite{JW28}. 
The bosonization technique in the Fock representation of the fields $\ph$ and $\phd$ 
has been applied for solving the Tomonaga-Luttinger (TL) model in \cite{Hald}-\cite{ML65}.  
The framework has been extended later \cite{Liguori:1999tw, Amaral:2005cqa} 
to the finite temperature Gibbs representation 
of $\ph$ and $\phd$. Both the Fock and Gibbs representations describe equilibrium physics. 
Our goal in what follows we will to apply the NESS representation, 
constructed in the previous section, for investigating 
the non-equilibrium TL liquid in the multi-terminal configuration shown in Fig. \ref{fig1}.

\subsection{The Tomonaga-Luttinger model on $\Gamma$}

The bulk dynamics is governed by the TL Lagrangian density 
\begin{equation}
{\cal L} = \ri \psi_1^*(\der_t - v_F\der_x)\psi_1 +  \ri \psi_2^*(\der_t + v_F\der_x)\psi_2\\ 
-g_+(\psi_1^* \psi_1+\psi_2^* \psi_2)^2 - g_-(\psi_1^* \psi_1-\psi_2^* \psi_2)^2 \, ,   
\label{lagr}
\end{equation}
where $\{\psi_\alpha (t,x,i)\,:\, \alpha =1,2\}$ 
are complex fermion fields, $v_F>0$ is the Fermi velocity and 
$g_\pm \in \RR$ are the coupling constants. 

The bulk theory has an obvious $U_L(1)\otimes U_R(1)$ symmetry. In fact, 
the Lagrangian density (\ref{lagr}) is left invariant by the two independent 
phase transformations 
\begin{equation}
\psi_\a  \rightarrow \e^{\ri s_\alpha } \psi_\a \, , \qquad \qquad \; \, 
\psi^*_\a \rightarrow \e^{-\ri s_\alpha } \psi^*_\a \, ,  \qquad s_\a \in \RR\, ,\quad  \a =1,2\, . 
\label{symm1}
\end{equation} 
implying the current conservation laws 
\begin{equation}
\der_t \rho_{Z} (t,x,i) -v_F\der_x j_{Z}  (t,x,i)= 0\, , 
\label{conservation}
\end{equation}
where the charge and current densities are given by   
\begin{equation}
\rho_{Z} (t,x,i) = 
\begin{cases}
[\psi^*_1\psi_1](t,x,i)\, , & Z=L\, , \\
[\psi^*_2\psi_2](t,x,i)\, , & Z=R\, ,
\end{cases} \qquad 
j_{Z} (t,x,i) = 
\begin{cases}
[\psi^*_1\psi_1](t,x,i)\, , & Z=L\, , \\
-[\psi^*_2\psi_2](t,x,i)\, , & Z=R\, . 
\end{cases} 
\label{lrcurrents}
\end{equation}
The currents $j_{Z}(t,x,i)$ have simple physical meaning: $j_{L}(t,x,i)$ and 
$j_{R}(t,x,i)$ represent the particle 
excitations moving along the edge $E_i$ towards and away of the vertex $V$ 
respectively. Interpreting the vertex as a 
defect, which can be characterized by some scattering matrix, 
the currents $j_{L}$ and $j_{R}$ describe therefore the {\it incoming} and {\it outgoing} flows.

\subsection{The current splitting boundary condition}

It is well known \cite{Hald}-\cite{ML65} that the 
TL model (\ref{lagr}) is exactly solvable on the line $\RR$.  
On the graph $\Gamma$ the situation is more involved, because one should take into account 
the boundary conditions in the vertex $V$. The conditions  
\begin{equation} 
\psi_1(t,0,i) = \sum_{j=1}^n \UU_{ij} \psi_2(t,0,j) \, , \qquad \UU\in U(n)\, , 
\label{bc2}
\end{equation}
which work in the free case $g_-=g_+=0$, do not lead \cite{nfll-99} to 
exactly solvable problem after switching on the TL interactions. The fact that on $\RR$ the 
quartic bulk interactions in (\ref{lagr}) are solved exactly 
via bosonization suggest to try boundary conditions which, differently from (\ref{bc2}), are 
formulated in terms of real boson fields. In this spirit and according our  
previous comments on the chiral currents (\ref{lrcurrents}), it is quite natural to consider  
\begin{equation}
j_R(t,0,i) = \sum_{k=1}^n \J_{ik}\, j_L(t,0,k) \, ,  \qquad \J \in O(n)\, , 
\label{bc3}
\end{equation} 
which has been proposed and explored first in the two-terminal case in \cite{SS}. An advantage of 
(\ref{bc3}) is the direct interpretation in terms of {\it gauge invariant physical observables}, 
which represent the basic building blocks of algebraic quantum field theory (see e.g. \cite{H}). 
In fact, (\ref{bc3}) describes the splitting in the 
vertex $V$ of the outgoing current $j_R(t,0,i)$ along the edge $E_i$ 
in incoming currents $j_L(t,0,k)$ along the edges $E_k$. For this reason we 
refer to $\J$ the as the {\it current splitting matrix} and show in 
the next subsection that $\J$ actually coincides with the boson scattering matrix $\S$. 

\subsection{Operator solution of the TL model on $\Gamma$} 

Referring for the details to \cite{Bellazzini:2008fu}, 
we recall here the anyon operator solution of the TL 
model on a star graph $\Gamma$. The solution provides a 
{\it unified description} of all anyon Luttinger liquids 
and is expressed in terms of the chiral fields (\ref{fil1},\ref{fil2}) 
and the parameters $\sigma,\, \tau \in \RR$ and the sound velocity $v\in \RR$ as follows:
\begin{eqnarray}  
\psi_1(t,x,i) &=&
\eta_i :\e^{\ri \sqrt {\pi} \left [\sigma \phr (vt-x) + \tau \phl (vt+x)\right ]}:\, , 
\label{psi1}\\
\psi_2(t,x,i) &=&
\eta_i:\e^{\ri \sqrt {\pi} \left [\tau \phr (vt-x) + \sigma \phl (vt+x)\right ]}:\, . 
\label{psi2}
\end{eqnarray}
Here $: \cdots :$ denotes the normal product in the RT algebra $\A$ and $\eta_i$ 
are some Klein factors, controlling the statistics of $\psi_\alpha $. In this respect 
we impose the general anyon exchange relation  
\begin{equation}
\psi_\alpha^*(t,x_1,i_1) \psi_\alpha (t,x_2,i_2) = 
\e^{(-1)^{\alpha} \ri \pi \k \eps(x_{12})} \psi_\alpha (t,x_2,i_2)\psi_\alpha^*(t,x_1,i_1)\, , 
\qquad x_1\not=x_2\, , 
\label{an1}
\end{equation} 
where $\eps(x)$ is the sign function and $\k > 0$ is the so called 
{\it statistical parameter} which interpolates between bosons ($\k$ - even integer) and 
fermions ($\k$ - odd integer). A simple realization of the Klein factors is  
\begin{equation} 
\eta_i = \frac{1}{\sqrt {2\pi}}:\e^{\pi\ri (\gamma_i + \gamma^*_i)}: \, , 
\label{KL1}
\end{equation}
where $\{\gamma_i,\, \gamma^*_i \, :\, i=1,...,n\}$ generate the auxiliary algebra 
\begin{equation} 
[\gamma_i\, ,\, \gamma_j] = [\gamma^*_i\, ,\, \gamma^*_j] = 0\, , \qquad 
[\gamma_i\, ,\, \gamma^*_j] = \ri \frac{\k}{2} \epsilon_{ij} \, , 
\label{KL2}
\end{equation} 
with $\epsilon_{ij}=-1$ for $i<j$, $\epsilon_{ii}=0$ and $\epsilon_{ij}=1$ for $i>j$. 

In order to fix the solution (\ref{psi1},\ref{psi2}) completely, one should 
determine the parameters $\sigma$, $\tau$ and $v$ in terms of coupling 
constants $g_\pm$ and the statistical parameter $\k$. 
Using a standard short distance expansion and (\ref{psi1},\ref{psi2}) 
one gets the charge and current densities\footnote{Without loss of generality 
we assume in what follows $\tau \geq 0$ and $\tau \not= \pm \sigma$.}  
\begin{equation}
\rho_\pm (t,x,i) \equiv (:\psi^*_1\psi_1: \pm  :\psi^*_2\psi_2:)(t,x,i) = 
\frac{-1}{2\sqrt {\pi }\zeta_\pm } 
\left [(\der \phr)(vt-x) \pm (\der \phl)(vt+x)\right ] , 
\label{rhopm}
\end{equation} 
\begin{equation}
j_\pm (t,x,i) = 
\frac{v}{2\sqrt {\pi }v_F \zeta_\pm} 
\left [(\der \phr)(vt-x) \mp (\der \phl)(vt+x)\right ]\, , 
\label{jpm}
\end{equation} 
where for convenience the variables 
\begin{equation}
\zeta_\pm =\tau\pm\sigma \, . 
\label{zpm}
\end{equation} 
have been introduced. The normalization of (\ref{rhopm}) 
is fixed \cite{Bellazzini:2008fu} by the Ward identities associated with 
the electric charge $Q_+$ and the helicity $Q_-$ defined by 
\begin{equation}
Q_\pm = \sum_{i=1}^n \int_0^\infty \rd x\, \rho_\pm(t,x,i)\, .  
\label{charges}
\end{equation}
The normalization of (\ref{jpm}) in turn is determined by the conservation law 
\begin{equation}
\der_t \rho_\pm (t,x,i) - v_F\der_x j_\pm (t,x,i) = 0\, .  
\label{vcons}
\end{equation}
Plugging (\ref{psi1},\ref{psi2},\ref{rhopm}) 
in the quantum equations of motion 
\begin{eqnarray}
\ri [\der_t +(-1)^\alpha v_F\der_x] \psi_\alpha (t,x,i)= \nonumber 
\qquad \qquad \qquad \qquad \\
2 [g_+:\rho_+(t,x,i)\psi_\alpha : (t,x,i) -(-1)^\alpha g_- :\rho_-(t,x,i)\psi_\alpha :(t,x,i)]\, , 
\label{TLeqm}
\end{eqnarray}
one finds 
\begin{eqnarray} 
v \zeta_+^2 &=& v_F \k +\frac{2}{\pi}g_+ \, , 
\label{sys2}\\
v \zeta_-^2 &=& v_F \k +\frac{2}{\pi}g_- \, .  
\label{sys3}
\end{eqnarray} 
Moreover, the exchange relation (\ref{an1}) implies 
\begin{equation}
\zeta_+\, \zeta_- = \kappa\, . 
\label{sys1} 
\end{equation} 
Eqs. (\ref{sys1},\ref{sys2},\ref{sys3}) provide a system for determining 
$v$ and $\zeta_\pm$ (or equivalently $\sigma$ and $\tau$) in terms of 
$\k$ and $g_\pm$. The solution is 
\begin{eqnarray}
\zeta_\pm^2 &=& \kappa
\left(\frac{\pi \kappa v_F+2g_+}{\pi \kappa v_F+2g_-}\right)^{\pm 1/2}\, , 
\label{z}\\ 
v&=&\frac{\sqrt{(\pi \kappa v_F+2g_-)(\pi \kappa v_F+2g_+)}}{\pi \kappa}\, , 
\label{v}
\end{eqnarray} 
where the positive roots are taken in the right hand side. 
The relations (\ref{z}) and (\ref{v}) represent the anyon 
generalization \cite{Bellazzini:2008fu}  
of the well known result for canonical fermions $\k=1$, where an alternative and frequently 
used notation \cite{voit-95} is  
\begin{equation}
g_2=2(g_+-g_-)\, , \qquad g_4=2(g_+ + g_-)\, , \qquad K=\zeta_-^2=\zeta_+^{-2}\, . 
\label{altnot}
\end{equation}

Considering the general anyon solution (\ref{z}, \ref{v}), we assume in 
what follows that the parameters $\{\k,\, g_\pm\}$ belong to the domain  
\begin{equation}
{\cal D} =\{\k>0,\; 2g_\pm> -\pi \kappa v_F\} \, ,  
\label{physcond}
\end{equation} 
which ensures that $\sigma$, $\tau$ and $v$ are real and finite. 

Let us discuss finally the current splitting boundary condition (\ref{bc3}) and establish 
the relation between $\J$ and $\S$. Expressing the chiral currents 
$j_Z$ in terms of the chiral fields $\phz$, one finds 
\begin{eqnarray}
j_R(t,x,i) &=& \frac{1}{2} (\zeta_-j_- + \zeta_+j_+)(t,x,i) = \frac{v}{2\sqrt {\pi }v_F} \der \phr (vt-x) \, , 
\label{qcc1} \\
j_L(t,x,i) &=& \frac{1}{2} (\zeta_-j_- - \zeta_+j_+)(t,x,i)= \frac{v}{2\sqrt {\pi }v_F} \der \phl (vt+x) \, ,
\label{qcc2}
\end{eqnarray}
which, according to (\ref{bc1}) satisfy the current splitting boundary condition (\ref{bc3}), provided that  
\begin{equation}
\J = \S \in O(n) \, . 
\label{bc4}
\end{equation}

The symmetry content of the TL junction is strongly influenced by (\ref{bc4}). The point is that 
in the presence of a defect the continuity equation (\ref{vcons}) alone is not enough to ensure the 
electric charge conservation. A direct computation shows indeed that 
\begin{equation}
\der_t Q_+ =  \frac{v}{2\sqrt {\pi} \zeta_+} \sum_{k=1}^n 
\left (1-\sum_{i=1}^n \S_{ik} \right )(\der \varphi_{k,L})(vt)\, . 
\label{sym1}
\end{equation} 
The independence of $\phl$ implies that 
the electric charge $Q_+$ is conserved if and only if  
\begin{equation} 
\sum_{i=1}^n \S_{ik} = 1\, , \qquad \forall\; k=1,...,n\, ,   
\label{K3}
\end{equation} 
which, as expected, is equivalent to the Kirchhoff rule 
\begin{equation} 
\sum_{i=1}^n j_+ (t,0,i) = 0\, . 
\label{K2}
\end{equation} 
Since $\S \in O(n)$, one infers from (\ref{K3}) that $\S^t$ satisfies (\ref{K3}) as well. 
Therefore, the electric charge $Q_+$ is conserved for those $\S$, whose entries along each column 
(line) sum up to 1. In geometric terms, these scattering matrices belong to the stability 
subgroup $O_{\bf v}\subset O(n)$ of the $n$-vector ${\bf v} = (1,1,...,1)$. An explicit parametrization 
of $O_{\bf v}$ in terms of angular variables is given in \cite{Bellazzini:2009nk}.  

Summarizing, the condition $\S\in O(n)$ guaranties the energy conservation in the TL-junction. 
Concerning the electric charge $Q_+$, one must distinguish two different regimes. $Q_+$ is 
conserved for $\S\in O_{\bf v}$. If instead $\S$ belongs to the complement $\tO_{\bf v} \equiv  
O(n)\setminus O_{\bf v}$, there is an external incoming or outgoing charge flow in the junction 
and $Q_+$ is not conserved. The possibility to describe such imperfect junctions is a remarkable 
feature of the current splitting boundary condition (\ref{bc3}). The physical details about 
the charge transport in the junction are discussed in section 4.2 below.

\subsection{NESS representation and chemical potentials} 

The crucial property of the operator solution 
(\ref{psi1},\ref{psi2},\ref{z},\ref{v}) is that it is 
universal, meaning that it applies for any representation of the 
chiral field algebra generated by $\phz$. The Fock and Gibbs 
representations have been largely studied and 
describe the equilibrium properties of the TL model on $\Gamma$. In order to explore 
the behavior of the Luttinger liquid away from equilibrium, we 
investigate below the operator solution in the NESS representation of the RT algebra 
$\A$, constructed in section 2.2. 

The first step in this direction is the introduction of the 
{\it fermion} chemical potentials 
\begin{equation}
\mu_i = k_F -V_i \, , 
\label{fcp1}
\end{equation}
where $k_F$ defines the Fermi energy for $\k=1$ and $V_i$ is the external voltage 
applied to the thermal reservoir in the edge $E_i$ of  
Fig. \ref{fig2}. In what follows we keep $k_F$ fixed and vary eventually the 
gate voltages $V_i$. As already mentioned, the {\it boson} chemical potential 
$\mu_b<0$ has been introduced for avoiding some infrared singularities at the boson 
level and has nothing to do with $\mu_i$. In fact, in the chiral correlators (\ref{ll}-\ref{rr}) 
we already performed the limit $\mu_b \to 0^-$. 
In order to recover $\mu_i$, following \cite{Liguori:1999tw} we introduce the shift $\alpha_\mu$, defined by  
\begin{equation}
\phl (\xi) \longmapsto (\alpha_\mu \phl)(\xi ) = \phl (\xi) -\frac{\xi}{\sqrt \pi\, \zeta_+}\, \mu_i 
\label{fcp2}
\end{equation}
and, consistently with the boundary condition (\ref{bc1}), 
\begin{equation}
\phr (\xi) \longmapsto (\alpha_\mu \phr)(\xi ) = \phr (\xi) -\frac{\xi}{\sqrt \pi\, \zeta_+}\sum_{j=1}^n \S_{ij} \mu_j \, . 
\label{fcp3}
\end{equation}
The transformations (\ref{fcp1},\ref{fcp2}) extend to an automorphism $\alpha_{\mu}$ 
on the whole algebra generated by the chiral fields $\phz$, which is 
directly implemented in the operator solution (\ref{psi1},\ref{psi2},\ref{rhopm},\ref{jpm}). 
At this stage the TL correlation functions in the NESS are defined by 
\begin{equation}
\langle \O_1[\ph_{i_1,Z}] \cdots \O_k[\ph_{i_k,Z}] \rangle_{\beta,\mu} = 
\langle \O_1[\alpha_\mu \ph_{i_1,Z}] \cdots 
\O_k[\alpha_\mu \ph_{i_k,Z}]\rangle_{\beta}\, . 
\label{gcf}
\end{equation} 

In the rest of the paper we focus on the correlation functions (\ref{gcf}), which capture 
the physical properties of the Luttinger liquid with the current splitting boundary 
condition (\ref{bc3}) away from equilibrium. We will show in particular that 
(\ref{gcf}) satisfy the Kubo-Martin-Schwinger (KMS) condition \cite{BR, H} at equilibrium, 
which justifies the introduction of the chemical potentials $\mu_i$ by 
means of (\ref{fcp2},\ref{fcp3}).

\section{Non-equilibrium TL correlation functions}
\bigskip 

\subsection{Anyon correlators}
\medskip 

We derive here the two-point correlators of $\psi_\alpha (t,x,i)$ defined by (\ref{psi1}, \ref{psi2}) 
in the NESS and discuss their properties. For this purpose we extend away from equilibrium 
the finite temperature results of \cite{Liguori:1999tw}. Using (\ref{gcf2}), for $\psi_1$ one finds 
\begin{eqnarray}
\langle \psi_1^*(t_1,x_1,i)\psi_1(t_2,x_2,j)\rangle_{\beta, \mu} = 
A_{ij}\, B_{ij}(t_{1,2},x_{1,2};\mu) \times
\qquad \qquad \qquad \qquad 
\nonumber \\
\left \{\frac{1}{\frac{\beta_i}{\pi} \sinh \left [\frac{\pi}{\beta_i}(vt_{12}+ \tx_{12}) -
\ri \varepsilon \right ]}\right \}^{\sigma \tau \S_{ij}^t} 
\left \{\frac{1}{\frac{\beta_j}{\pi} \sinh \left [\frac{\pi}{\beta_j}(vt_{12} - \tx_{12}) -
\ri \varepsilon \right ]}\right \}^{\sigma \tau \S_{ij}} \quad \; \; 
\nonumber \\
\left \{\frac{1}{\frac{\beta_i}{\pi} \sinh \left [\frac{\pi}{\beta_i}(vt_{12}+ x_{12}) -
\ri \varepsilon \right ]}\right \}^{\tau^2 \delta_{ij}} 
\prod_{k=1}^n
\left \{\frac{1}{\frac{\beta_k}{\pi} \sinh \left [\frac{\pi}{\beta_k}(vt_{12} - x_{12}) -
\ri \varepsilon \right ]}\right \}^{\sigma^2 \S_{ik}\S_{kj}^t} \, ,
\label{anc1}
\end{eqnarray} 
where 
\begin{equation}
A_{ij} = \frac{\e^{\ri \pi^2 \kappa \epsilon_{ij}/2}}{2\pi} 
\left (\frac{1}{2\ri}\right )^{(\sigma^2 +\tau^2)\delta_{ij} +\sigma \tau (\S_{ij}+\S^t_{ij})} \, , 
\label{anc2}
\end{equation}
\begin{equation}
B_{ij}(t_{1,2},x_{1,2};\mu) = \e^{\ri \{\tau [(vt_1+x_1)\mu_i-(vt_2+x_2)\mu_j]+
\sigma[(vt_1-x_1)\sum_{k=1}^n\S_{ik}\mu_k -(vt_2-x_2)\sum_{k=1}^n\S_{jk}\mu_k]\}/(\sigma+\tau)} \, .
\label{anc4}
\end{equation}
The $\psi_2$-correlator has the analogous form,  
\begin{equation}
\langle \psi_2^*(t_1,x_1,i)\psi_2(t_2,x_2,j)\rangle_{\beta, \mu} = (\ref{anc1})\quad {\rm with}\quad \sigma 
\leftrightarrow \tau \, . 
\label{anc22}
\end{equation}

The TL junction involves two types of $\psi_1$-$\psi_2$ interactions. First, the Lagrangian (\ref{lagr}) 
contains a $\psi_1$-$\psi_2$ {\it bulk coupling} proportional to $(g_+-g_-)$. Second, the current splitting 
boundary condition (\ref{bc3}) provides an additional {\it boundary interaction} described by the mixed 
left-right correlators (\ref{lr}, \ref{rl})). Consequently, the mixed $\psi_1$-$\psi_2$ correlators are non-trivial and 
have the form 
\begin{eqnarray}
\langle \psi_1^*(t_1,x_1,i)\psi_2(t_2,x_2,j)\rangle_{\beta, \mu} = 
\tA_{ij}\, \tB_{ij}(t_{1,2},x_{1,2};\mu) \times
\qquad \qquad \qquad \qquad 
\nonumber \\
\left \{\frac{1}{\frac{\beta_i}{\pi} \sinh \left [\frac{\pi}{\beta_i}(vt_{12}+ \tx_{12}) -
\ri \varepsilon \right ]}\right \}^{\tau^2 \S_{ij}^t} 
\left \{\frac{1}{\frac{\beta_j}{\pi} \sinh \left [\frac{\pi}{\beta_j}(vt_{12} - \tx_{12}) -
\ri \varepsilon \right ]}\right \}^{\sigma^2 \S_{ij}} \quad \; \; 
\nonumber \\
\left \{\frac{1}{\frac{\beta_i}{\pi} \sinh \left [\frac{\pi}{\beta_i}(vt_{12}+ x_{12}) -
\ri \varepsilon \right ]}\right \}^{\sigma \tau \delta_{ij}} 
\prod_{k=1}^n
\left \{\frac{1}{\frac{\beta_k}{\pi} \sinh \left [\frac{\pi}{\beta_k}(vt_{12} - x_{12}) -
\ri \varepsilon \right ]}\right \}^{\sigma \tau \S_{ik}\S_{kj}^t} \, ,
\label{manc1}
\end{eqnarray} 
where 
\begin{equation}
\tA_{ij} = \frac{\e^{\ri \pi^2 \kappa \epsilon_{ij}/2}}{2\pi} 
\left (\frac{1}{2\ri}\right )^{\sigma^2 \S_{ij} +\tau^2 \S^t_{ij} +2\sigma \tau \delta_{ij}} \, , 
\label{manc2}
\end{equation}
\begin{equation}
\tB_{ij}(t_{1,2},x_{1,2};\mu) = \e^{\ri \{\tau [(vt_1+x_1)\mu_i-(vt_2-x_2)\sum_{k=1}^n\S_{jk}\mu_k]+
\sigma[(vt_1-x_1)\sum_{k=1}^n\S_{ik}\mu_k-(vt_2+x_2)\mu_j ]\}/(\sigma+\tau)} \, .
\label{manc4}
\end{equation}
Finally,  
\begin{equation}
\langle \psi_2^*(t_1,x_1,i)\psi_1(t_2,x_2,j)\rangle_{\beta, \mu} = (\ref{manc1})\quad {\rm with}\quad \sigma 
\leftrightarrow \tau \, . 
\label{manc22}
\end{equation}

As expected, in the equilibrium limit $\beta_i = \beta$ and $\mu_i=\mu$ for all $i$, the correlators 
(\ref{anc1})--(\ref{manc22}) simplify and satisfy the KMS condition, 
which represents a non-trivial check both on the computation and on the 
shift (\ref{fcp2}, \ref{fcp3}) introducing the chemical potentials. 
Let us consider for instance (\ref{anc1}), which in this limit 
takes the form 
\begin{eqnarray}
\langle \psi_1^*(t_1,x_1,i)\psi_1(t_2,x_2,j)\rangle_{\beta, \mu} = A_{ij}\, 
\e^{\ri [\tau (vt_{12}+x_{12}) + \sigma (vt_{12}-x_{12})]\mu /(\sigma+\tau)}\times
\qquad \qquad 
\nonumber \\
\left \{\frac{1}{\frac{\beta }{\pi} \sinh \left [\frac{\pi}{\beta }(vt_{12}+ \tx_{12}) -
\ri \varepsilon \right ]}\right \}^{\sigma \tau \S_{ij}^t} 
\left \{\frac{1}{\frac{\beta }{\pi} \sinh \left [\frac{\pi}{\beta }(vt_{12} - \tx_{12}) -
\ri \varepsilon \right ]}\right \}^{\sigma \tau \S_{ij}} \quad \; \; 
\nonumber \\
\left \{\frac{1}{\frac{\beta }{\pi} \sinh \left [\frac{\pi}{\beta }(vt_{12}+ x_{12}) -
\ri \varepsilon \right ]}\right \}^{\tau^2 \delta_{ij}} 
\left \{\frac{1}{\frac{\beta }{\pi} \sinh \left [\frac{\pi}{\beta }(vt_{12} - x_{12}) -
\ri \varepsilon \right ]}\right \}^{\sigma^2 \delta_{ij}} \, . \quad 
\label{anc5}
\end{eqnarray} 
Recalling that the KMS automorphism $\varrho_s$ acts on $\psi_\alpha$ as follows, 
\begin{equation}
\left [\varrho_s \psi_\alpha\right ] (t,x,i) = \e^{\ri s \mu}\, \psi_\alpha (t + s/v,x,i)\, , 
\label{anc6}
\end{equation}
one can check that the equilibrium correlator (\ref{anc5}) satisfies the KMS condition 
\begin{equation}
\langle \psi_1^*(t_1,x_1,i)\left [\varrho_{s+\ri \beta}\psi_1\right ](t_2,x_2,j)\rangle_{\beta, \mu} 
= \langle \left [\varrho_{s}\psi_1\right ](t_2,x_2,j)\psi_1^*(t_1,x_1,i)\rangle_{\beta, \mu}  
\label{anc7}
\end{equation} 
for all values of the statistical parameter $\k$. 

The critical scaling dimensions $d_i$ can be extracted from (\ref{anc1})--(\ref{manc22}) in the limit 
$\beta_i \to \infty$ and $\mu_i \to 0$. Because of the operator mixing, this is a subtle issue, which 
has been discussed in full detail in \cite{Bellazzini:2009nk}. One gets, 
\begin{equation} 
d_i = \frac{1}{2}(\sigma^2 + \tau^2) + \sigma \tau s_i \, , \qquad i=1,...,n \, ,  
\label{dimensions0}
\end{equation} 
where $s_i=\pm 1$ are the eigenvalues of $\S$. As already observed in \cite{Bellazzini:2006kh}, 
the impact of the vertex interaction is captured 
by the term $\sigma \tau s_i$, which preserves unitarity in the sense of conformal field theory 
because $d_i \geq 0$. 

A remarkable special case is obtained by setting $g_+=g_-\equiv g$. In this case the bulk $\psi_1$-$\psi_2$ 
coupling vanishes and one is left only with the boundary interaction induced by the 
current splitting boundary condition (\ref{bc3}). From (\ref{v}) one gets 
\begin{equation}
v=v_F + \frac{2g}{\pi \kappa }\, , 
\label{nv}
\end{equation} 
and (using (\ref{sys1}) with $\k > 0$ and $\tau \geq 0$)   
\begin{equation}
\sigma =0\, , \quad \tau = \sqrt {\kappa }\, .
\label{nst}
\end{equation}
Inserting (\ref{nst}) in (\ref{anc1})--(\ref{anc22}) and localizing the fields in the same 
edge (i.e. setting $i=j$), one finds that the correlation functions simplify to  
\begin{eqnarray}
C_{11}(vt_{12}+x_{12},i;\beta,\mu) \equiv 
\langle \psi_1^*(t_1,x_1,i)\psi_1(t_2,x_2,i)\rangle_{\beta, \mu} = 
\nonumber \\
\frac{1}{2\pi} 
\left (\frac{1}{2\ri}\right )^\k 
\e^{\ri (vt_{12}+ x_{12})\mu_i}
\left \{\frac{1}{\frac{\beta_i}{\pi} \sinh \left [\frac{\pi}{\beta_i}(vt_{12}+ x_{12}) -
\ri \varepsilon \right ]}\right \}^{\k}\, ,
\label{panc1}
\end{eqnarray} 
\begin{eqnarray}
C_{22}(vt_{12}-x_{12},i;\beta,\mu) \equiv 
\langle \psi_2^*(t_1,x_1,i)\psi_2(t_2,x_2,i)\rangle_{\beta, \mu} = 
\qquad \quad 
\nonumber \\
\frac{1}{2\pi} 
\left (\frac{1}{2\ri}\right )^\k 
\prod_{k=1}^n \e^{\ri (vt_{12}- x_{12})\S_{ik}\mu_k} 
\left \{\frac{1}{\frac{\beta_k}{\pi} \sinh \left [\frac{\pi}{\beta_k}(vt_{12}- x_{12}) -
\ri \varepsilon \right ]}\right \}^{\k \S_{ik}^2}. 
\label{panc2}
\end{eqnarray} 
The condition $g_+=g_-$ and the left-right asymmetry of the NESS construction in section 2 
imply that only left moving (incoming) excitations contribute to $C_{11}$, which therefore  
coincides with the equilibrium correlator \cite{Liguori:1999tw}. 
All the non-equilibrium features are captured by $C_{22}$, which 
involves only right moving (outgoing) excitations. In fact, in spite of being 
localized in the edge $E_i$ of the graph, (\ref{panc2}) depends on the temperatures and 
chemical potentials of the all $n$ edges. 

It is instructive for this reason to derive and compare 
the Fourier transforms of (\ref{panc1}, \ref{panc2}). We will show first that they can be expressed in terms 
of the finite temperature TL {\it anyon distribution} discovered in \cite{Liguori:1999tw}. Consider in fact 
\begin{equation}
\hC_{11}(E,p,i;\beta,\mu) \equiv \int_{-\infty}^\infty \rd t 
\int_{-\infty}^\infty \rd x\, \e^{-\ri (Evt +px)}\, C_{11}(vt +x,i;\beta,\mu)\, . 
\label{ad1}
\end{equation}
Plugging (\ref{panc1}) in (\ref{ad1}) one gets 
\begin{equation}
\hC_{11}(E,p,i;\beta,\mu) = \frac{\pi^\kappa}{v} \delta(E-p)\, d(p-\mu_i,\beta_i;\k) \, , 
\label{ad2}
\end{equation}
where the $\delta$-function fixes the dispersion relation and 
$d$ is the equilibrium anyon momentum distribution \cite{Liguori:1999tw}
\begin{eqnarray}
d(p,\beta;\k) = \frac{\beta^{1-\k} \e^{-\beta p/2}}{2\pi}\, B \left (\frac{\k}{2} - \frac{\ri}{2\pi} \beta p \, , 
\frac{\k}{2} + \frac{\ri}{2\pi} \beta p\right ) = \qquad \qquad \qquad 
\nonumber \\
\frac{\beta^{1-\k} \e^{-\beta p/2}}{2\pi \Gamma (\k)}\,   
\Gamma \left (\frac{\k}{2} - \frac{\ri}{2\pi} \beta p\right ) 
\Gamma \left (\frac{\k}{2} + \frac{\ri}{2\pi} \beta p\right ) \, , \qquad \k>0\, , 
\label{ad3}
\end{eqnarray}
$B$ and $\Gamma$ being the beta and gamma functions (Euler's integrals of first and second kind respectively). 
Notice that for $\k\not=1$ the distribution (\ref{ad3}) depends on both $\beta$ and $p$ and not only 
on the dimensionless combination $\beta p$. Eq. (\ref{ad3}) defines a smooth function of 
$p\in \RR$, which satisfies
\begin{equation} 
\lim_{p\to 0}d(p,\beta;\k) = 
\frac{\beta^{1-\k} \Gamma^2(\k/2)}{2\pi \Gamma (\k)}\, , 
\label{as0}
\end{equation}  
and has the following asymptotic behavior: 
\begin{eqnarray}
\lim_{p\to \infty}d(p,\beta;\k) &=& 0\, , \quad \forall\;  \k>0\, , 
\label{as1}\\
\lim_{p\to -\infty}d(p,\beta;\k) &=& 
\begin{cases}
0\, , \quad &0<\k<1\, ,\\
1\, , \quad &\k=1\, ,\\
\infty \, , \quad &\k >1\, . 
\end{cases}
\label{as2}
\end{eqnarray}
{}For positive integer $\k$ (i.e. for fermions and bosons) the distribution (\ref{ad3}) 
simplifies to 
\begin{equation} 
d(p,\beta;\k) = 
\begin{cases}
\frac{1}{(1+\e^{\beta p})} \frac{\beta^{-2(n-1)}}{[2(n-1)]!} \prod_{j=1}^{n-1}
\bigl |j - \frac{1}{2} - \frac{\ri}{2\pi} p\beta \bigr |^2\, , & \k = 2n-1 \, , \\
\\
\frac{1}{(\e^{\beta p}-1)} \frac{2\pi \beta^{-2n} }{(2n-1)! p} \prod_{j=0}^{n-1}
\bigl |j- \frac{\ri}{2\pi} p\beta \bigr |^2\, , & \k = 2n\, , 
\end{cases} 
\label{intad}
\end{equation} 
where $n=1,2,...$ and the familiar Fermi and Bose distributions appear as prefactors. 
The first two fermion and boson distributions are  
\begin{eqnarray}
d(p, \beta;1) = \frac{1}{(1+\e^{\beta p})}\, , \qquad \qquad 
d(p, \beta; 3) = \frac{1}{(1+\e^{\beta p})} \frac{(\pi^2+p^2\beta^2)}{8\pi^2\beta^2}\, , 
\label{ad4f} \\
d(p, \beta; 2) = \frac{1}{(\e^{\beta p}-1)} \frac{p}{2\pi}\, , \qquad 
d(p, \beta; 4) = \frac{1}{(\e^{\beta p}-1)} \frac{p(4\pi^2+p^2\beta^2)}{48\pi^3\beta^2}\, .
\label{ad4b}
\end{eqnarray}
As expected, in the fermion point $\k=1$ of the TL liquid one gets 
the familiar Fermi distribution. In spite of the fact that the remaining boson 
and fermion points ($\k=2,3,...$) have been established in \cite{Liguori:1999tw, Ilieva:2000cj}
more then a decade ago, to our knowledge their physical meaning and 
potential applications of (\ref{intad}) have not been fully explored. 

\begin{figure}[h]
\begin{center}
\begin{picture}(500,110)(-120,20) 
\includegraphics[scale=0.8]{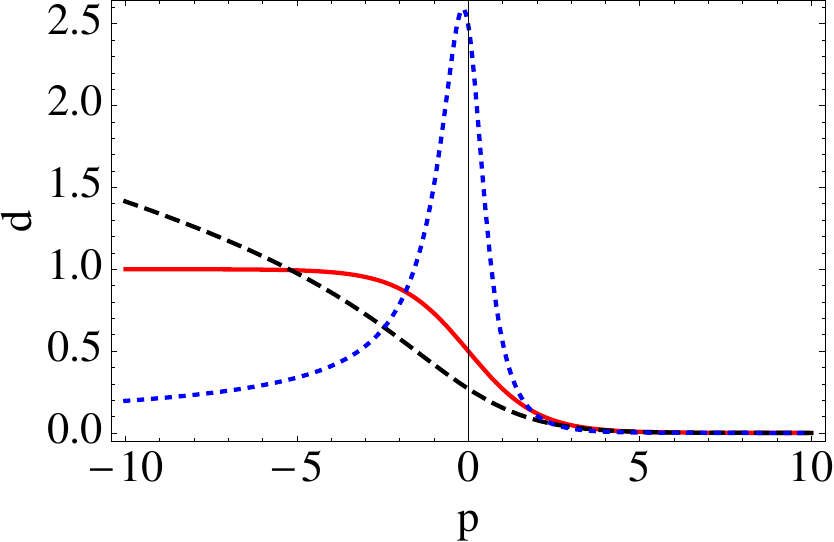}
\end{picture} 
\end{center}
\caption{The distribution $d$ at fixed temperature $\beta=1$ for $\k=1/2$ (dotted blue line) and 
$\k=3/2$ (dashed black line) compared 
to the Fermi distribution $\k=1$ (continuous red curve).}
\label{fig3}
\end{figure} 

\noindent 
\begin{figure}[h]
\begin{center}
\begin{picture}(500,110)(-120,20) 
\includegraphics[scale=0.8]{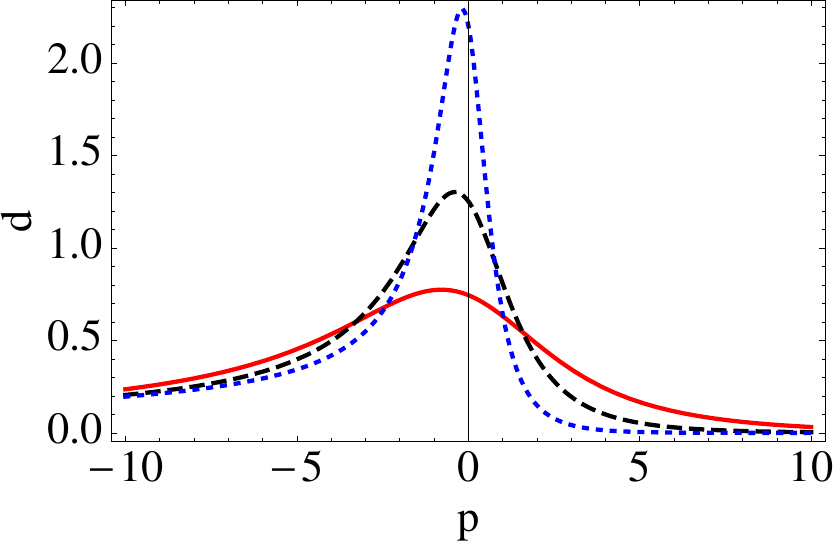}
\end{picture} 
\end{center}
\caption{The distribution $d$ at fixed $\k=1/4$ for different temperatures $\beta =0.2$ (continuos red line), 
$\beta=0.4$ (dashed black line) and $\beta=0.8$ (dotted blue line).}
\label{fig4}
\end{figure} 

In order to give an idea about the anyon distributions in the interval $0<\k< 1$, we show  
some of them in Fig.\ref{fig3}, where the standard Fermi distribution (continuous red curve) 
is given for comparison. Fig.\ref{fig4} displays the behavior of the 
anyon distribution (\ref{ad3}) for fixed $\k=1/4$ and different 
temperatures. For $0<\k<1$ and with decreasing of the temperature $T\sim 1/\beta$ one observes 
the formation of a sharp peak at $p=0$ (in agreement with eq. (\ref{as0})), which signals a 
condensation-like phenomenon \cite{Liguori:1999tw}.  

Concerning the Fourier transform of (\ref{panc2}), it is useful to consider first the 
case when all the temperatures are equal ($\beta_i = \beta$), the system being 
driven away from equilibrium only by the voltages $V_i$. In this case 
\begin{eqnarray}
C_{22}(vt_{12}-x_{12},i;\beta,\mu) \equiv 
\langle \psi_2^*(t_1,x_1,i)\psi_2(t_2,x_2,i)\rangle_{\beta, \mu} = 
\qquad \quad 
\nonumber \\
\frac{1}{2\pi} 
\left (\frac{1}{2\ri}\right )^\k 
\e^{\ri (vt_{12}- x_{12})\sum_{k=1}^n\S_{ik}\mu_k} 
\left \{\frac{1}{\frac{\beta}{\pi} \sinh \left [\frac{\pi}{\beta }(vt_{12}- x_{12}) -
\ri \varepsilon \right ]}\right \}^{\k}. 
\label{panc22}
\end{eqnarray}  
and therefore 
\begin{equation}
\hC_{22}(E,p,i:\beta,\mu) = \frac{\pi^\kappa}{v} \delta(E-p)\, d(p-\sum_{k=1}^n\S_{ik}\mu_k,\beta;\k)\, . 
\label{ad22}
\end{equation}
One has still the equilibrium distribution, with the energy shifted by a linear 
combination of the chemical potentials $\mu_k$, whose coefficients are the $\S$-matrix elements. 

Finally, in the coordinate space the general expression (\ref{panc2}) is a product of 
$C_{11}$-factors with different temperatures and chemical potentials. 
One gets therefore in momentum space the nested convolution formula
\begin{eqnarray}
\hC_{22}(E,p,i:\beta,\mu) = \frac{\pi^\kappa}{v} \delta(E-p)\, 
\int_{-\infty}^\infty \frac{\rd k_1}{2\pi} \int_{-\infty}^\infty \frac{\rd k_2}{2\pi} 
\cdots \int_{-\infty}^\infty \frac{\rd k_{n-1}}{2\pi} \times \qquad 
\nonumber \\
d(k_1-\S_{i1}\mu_1,\beta_1;\k\S^2_{i1})d(k_2-\S_{i2}\mu_2-k_1,\beta_2;\k\S^2_{i2})
\cdots d(p-\S_{in}\mu_n-k_{n-1},\beta_n;\k\S^2_{in})\, .  
\nonumber \\ 
\label{ad223}
\end{eqnarray} 
Being a convolution of distributions, (\ref{ad223}) is also a well defined distribution. The NESS 
$\Omega_{\beta, \mu}$ has therefore a remarkable property: the associated {\it non-equilibrium  
distribution} is simply a {\it convolution of equilibrium distributions} with different 
temperatures and chemical potentials. 

\begin{figure}[h]
\begin{center}
\begin{picture}(500,110)(-120,20) 
\includegraphics[scale=0.8]{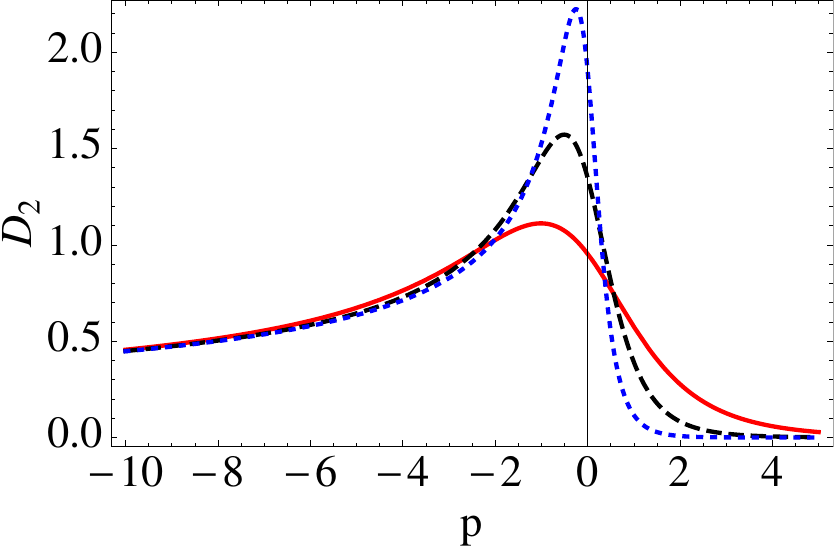}
\end{picture} 
\end{center}
\caption{The distribution $D_2$ at fixed $\k=1/2$ and $\theta =\pi/4$ for different temperatures 
$(\beta_1,\beta_2) =(1/2,1)$ (continuos red line), 
$(\beta_1,\beta_2) =(1,2)$ (dashed black line) and $(\beta_1,\beta_2) =(2,4)$ (dotted blue line).}
\label{fig5}
\end{figure}

\begin{figure}[h]
\begin{center}
\begin{picture}(500,110)(-120,20) 
\includegraphics[scale=0.8]{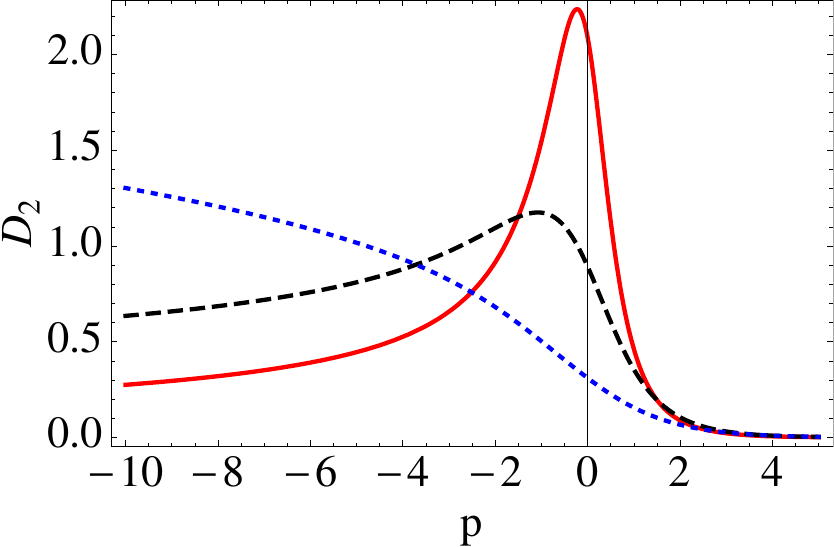}
\end{picture} 
\end{center}
\caption{The distribution $D_2$ at fixed $\beta_1=1$, $\beta_2=6$ and $\theta=\pi/6$ 
for statistical parameters $\k =1/3$ (continuos red line), 
$\k=2/3$ (dashed black line) and $\k=4/3$ (dotted blue line).}
\label{fig6}
\end{figure}

Since the general form of (\ref{ad223}) is quite complicated, it is instructive to consider below the 
case $n=2$ and $\mu_1=\mu_2=0$, focusing on 
\begin{equation}
D_2(p;\beta_1,\beta_2;\k,\theta) = 
\int_{-\infty}^\infty \frac{\rd k}{2\pi}\, d(k,\beta_1;\k \cos^2 \theta )d(p-k,\beta_2;\k \sin^2 \theta)\, , 
\qquad \theta \in [0,\pi)\, .  
\label{b1}
\end{equation}
Using the $x$-space representation, at equal temperatures one finds the relation 
\begin{equation} 
D_2(p;\beta,\beta;\k,\theta) = d(k,\beta ;\k)\, , \qquad \forall\; \theta \in [0,\pi)\, .  
\label{b2}
\end{equation} 
{}For $\beta_1 \not=\beta_2$ the convolution $D_2$ defines a new distribution. Since we were 
not able to determine its explicit analytic form, we give some plots which are obtained numerically. 
The plots in Figs. \ref{fig5} and \ref{fig6} illustrate the behavior of $D_2$ for different values of 
$\beta_{1,2}$, $\k$ and $\theta$. We see that even for $\beta_1 \not=\beta_2$ the distribution $D_2$ is similar to $d$ with a kind of ``effective" temperature and statistical 
parameter depending on $\beta_{1,2}$, $\k$ and $\theta$. 

\noindent 

Summarizing, we derived above the two-point TL anyon correlation functions away from equilibrium. 
The results (\ref{anc1},\ref{manc1}) are expressed as products of $\S$-dependent powers of equilibrium 
correlators at different temperatures. In agreement with this fact the momentum space anyon NESS 
distribution is the convolution of equilibrium anyon distributions (\ref{ad223}). The equilibrium limit satisfies the KMS conditions. At criticality one is dealing with 
a $c=1$ conformal field theory, whose anomalous dimensions (\ref{dimensions0}) depend not only on the 
coupling constants $g_\pm$, but also on the scattering matrix $\S$. 
The above technique allows to compute higher anyon 
correlation functions as well, but in order to investigate the transport properties of the system, 
we concentrate below on the electric and energy current correlators away from equilibrium.

\subsection{Charge and heat transport}

\begin{figure}[h]
\begin{center}
\begin{picture}(500,110)(-120,20) 
\includegraphics[scale=0.8]{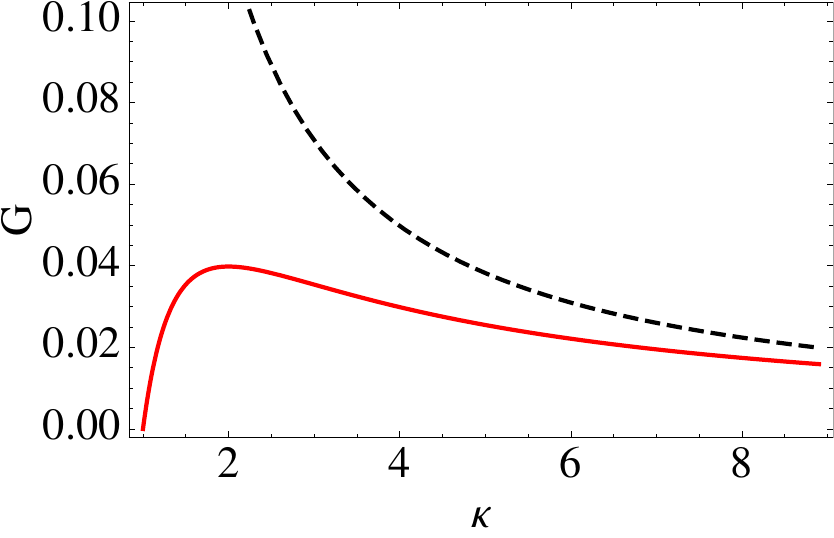}
\end{picture} 
\end{center}
\caption{Behavior of $G$ at $g_-=1$ (dashed black line) and  
$g_-=-1$ (continuous red line).}
\label{fig7}
\end{figure}

The charge transport in the NESS is described by 
\begin{eqnarray}
\langle \der_t Q_+ \rangle_{\beta, \mu} &=& \frac{v}{2\pi \zeta^2_+} \sum_{i,j=1}^n 
(\S_{ij} - \delta_{ij}) \mu_j\, , 
\label{ct00}\\
\langle j_+(t,x,i)\rangle_{\beta, \mu} &=& \frac{v}{2\pi v_F \zeta^2_+} 
\sum_{j=1}^n (\delta_{ij}-\S_{ij})\mu_j \, , 
\label{ct01}
\end{eqnarray} 
which follow by substituting (\ref{jpm},\ref{sym1}) in (\ref{gcf}). Eq. (\ref{ct00}) describes the 
external charge flow in the junction: it is constant in time and is incoming for 
$\langle \der_t Q_+ \rangle_{\beta, \mu} >0$ 
and outgoing for $\langle \der_t Q_+ \rangle_{\beta, \mu} <0$. Eq. (\ref{ct01}) determines 
instead the value of the currents along the leads. The charge balance 
\begin{equation}
\langle \der_t Q_+ \rangle_{\beta, \mu} +
v_F \sum_{i=1}^n\langle j_+(t,x,i)\rangle_{\beta, \mu} = 0  
\label{ct02}
\end{equation}
is satisfied and represents an useful check. If $\S\in O_{\bf v}$, the electric charge is 
conserved $\langle \der_t Q_+ \rangle_{\beta, \mu} =0$ and the $k_F$-dependence in  
(\ref{ct01}) drops out, leading to  
\begin{equation}
\langle j_+(t,x,i)\rangle_{\beta, \mu} = \frac{v}{2\pi v_F \zeta^2_+} 
\sum_{j=1}^n (\S_{ij}-\delta_{ij})V_j \, . 
\label{ct1}
\end{equation} 
The current (\ref{ct1}) satisfies the Kirchhoff rule (\ref{K2}) and vanishes at equilibrium ($V_i=V$ for all $i$) 
as it should be. The dependence on the statistical parameter $\k$ is explicit and deserves a comment. 
In the physical domain $\cal D$, defined by (\ref{physcond}), 
the overall coefficient in front of the sum in (\ref{ct1}) is positive, 
\begin{equation}
G(g_-,\k)\equiv \frac{v}{2\pi v_F\zeta_+^2} = \frac{\pi \kappa v_F + 2g_-}{2 \pi^2 \kappa^2 v_F} >0\, . 
\label{ct3}
\end{equation}
{}For $g_->0$ the coefficient $G$ decreases monotonically with $\k>0$. For 
$g_-<0$ one has that $\k>-2g_-/\pi v_F$ in the physical domain $\cal D$. In this case the 
coefficient $G$ increases in the interval $-2g_-/\pi v_F<\k < -4g_-/\pi v_F$, 
reaching the maximal value $-v_F/16g_-$ and decreases for $\k>-4g_-/\pi v_f$.  
This behavior is illustrated in Fig. \ref{fig7}. 

The current (\ref{ct1}) is proportional to the applied external voltages. Non-linear effects 
are absent in the critical regime under consideration, which implies the conductance tensor 
\begin{equation} 
\G_{ij} =G(g_-,\k)\sum_{j=1}^n (\S_{ij}-\delta_{ij})\, . 
\label{ct2}
\end{equation}
We see that the NESS approach, adopted in this paper, confirms the 
result for $\G$, obtained previously for $\k=1$ by different methods, 
including renormalization group techniques \cite{lrs-02, emabms-05}, 
linear response theory \cite{Bellazzini:2006kh, Bellazzini:2008mn} 
and conformal field theory \cite{coa-03, hc-08}. The novelty in (\ref{ct2}) 
is the explicit dependence on the statistical parameter $\k$, shown in Fig. \ref{fig7}. 

A similar computation gives the energy (heat) flow 
\begin{equation}
\langle \theta_{xt}(t,x,i)\rangle_{\beta,\mu} = 
\frac{v^2}{8\pi \zeta_+^2}\left [ \mu_i^2 - \left (\sum_{j=1}^n \S_{ij}\mu_j \right )^2\right ] + 
\frac{\pi v^2}{12}\sum_{j=1}^n \left (\delta_{ij} - \S_{ij}^2 \right )\frac{1}{\beta_j^2}\, , 
\label{ct04}
\end{equation} 
which satisfies the Kirchhoff rule (\ref{K1}) for all $\S \in O(n)$. For $\S\in O_{\bf v}$ 
the expression (\ref{ct04}) takes the form  
\begin{eqnarray}
\langle \theta_{xt}(t,x,i)\rangle_{\beta,\mu} = 
\qquad \qquad \qquad \qquad \qquad \qquad \nonumber \\
\frac{v^2}{8\pi \zeta_+^2}\left [ V_i^2 - 2k_F \sum_{j=1}^n (\delta_{ij}-\S_{ij})V_j - 
\left (\sum_{j=1}^n \S_{ij}V_j \right )^2\right ] + 
\frac{\pi v^2}{12}\sum_{j=1}^n \left (\delta_{ij} - \S_{ij}^2 \right )\frac{1}{\beta_j^2}\, . 
\label{ct4}
\end{eqnarray} 
The heat flow depends therefore not only on $k_F$ and the voltages $V_i$, but also on 
the temperatures $\beta_i$. 

As before, we consider for illustration the case of $n=2$ wires. Inserting the scattering matrices 
(\ref{ex1}) in (\ref{ct01}) one has 
\begin{equation}
\langle \der_t Q_+ \rangle_{\beta, \mu} =
\begin{cases} 
Gv_F[(\mu_1 +\mu_2)(\cos \theta-1)-(\mu_1-\mu_2)\sin \theta ]\, ,   & \; \;   {\rm det}\, \S =1\, , \\
Gv_F[(\mu_1 +\mu_2)(\sin \theta -1)+(\mu_1-\mu_2)\cos \theta ]\, ,   & \; \;   {\rm det}\, \S =-1\, . \\
\end{cases}
\label{ex2}
\end{equation} 
Therefore, 
\begin{equation}
\langle \der_t Q_+ \rangle_{\beta, \mu} = 0 \Longrightarrow 
\begin{cases} 
\theta =0 \quad \Longrightarrow j_+(t,x,1) = j_+(t,x,2) =0 \, ,   & \; \;   {\rm det}\, \S =1\, , \\
\theta =\pi/2 \Longrightarrow  j_+(t,x,1) = -j_+(t,x,2)=G(\mu_1-\mu_2)\, ,   & \; \;   {\rm det}\, \S =-1\, , \\
\end{cases}
\label{ex3}
\end{equation} 
corresponding respectively to full reflection (disconnected edges) and complete transmission in the junction. 
Finally, 
\begin{eqnarray}
\langle \theta_{xt}(t,x,1)\rangle_{\beta,\mu} = -\langle \theta_{xt}(t,x,2)\rangle_{\beta,\mu}=
\qquad \qquad \qquad \nonumber \\ 
\frac{v^2}{8\pi \zeta_+^2}[(\mu_1^2-\mu_2^2)\sin^2 \theta  -\mu_1\mu_2 \sin 2\theta ] + 
\frac{\pi v^2\sin^2 \theta}{12}\left (\frac{1}{\beta_1^2}-\frac{1}{\beta_2^2}\right ) \, ,  
\label{ex4}
\end{eqnarray} 
for both families in (\ref{ex1}).

\subsection{Quantum noise}

In this section we derive the {\it noise power} in the TL  
junction in Fig. \ref{fig1}. For this purpose we need \cite{bb-00} 
the two-point {\it connected} current-current correlator  
\begin{eqnarray}
\langle j_+(t_1,x_1,i) j_+(t_2,x_2,j) \rangle_{\beta, \mu}^{\rm conn} \equiv 
\qquad \qquad \qquad \qquad 
\nonumber \\
\langle j_+(t_1,x_1,i) j_+(t_2,x_2,j) \rangle_{\beta, \mu} - 
\langle j_+(t_1,x_1,i)\rangle_{\beta, \mu}\langle j_+(t_2,x_2,j) \rangle_{\beta, \mu}\, . 
\label{N1}
\end{eqnarray} 
After some algebra one finds 
\begin{eqnarray}
\langle j_+(t_1,x_1,i) j_+(t_2,x_2,j) \rangle_{\beta, \mu}^{\rm conn} \equiv 
\qquad \qquad \qquad \qquad 
\nonumber \\
\left (\frac{v}{2 v_f \zeta_+}\right )^2 
\Biggl \{\frac{1}{\beta_i^2 \sinh^2\left [\frac{\pi}{\beta_i}(vt_{12}+ x_{12}) -\ri \varepsilon \right ]} \delta_{ij} +
\sum_{l=1}^n \S_{il} \frac{1}{\beta_l^2 \sinh^2\left [\frac{\pi}{\beta_l}(vt_{12}- x_{12}) -
\ri \varepsilon \right ]} \S_{lj}^t
\quad \nonumber \\
 -\S_{ij} \frac{1}{\beta_j^2 \sinh^2\left [\frac{\pi}{\beta_j}(vt_{12}-\tx_{12}) -\ri \varepsilon \right ]}
- \frac{1}{\beta_i^2 \sinh^2\left [\frac{\pi}{\beta_i} (vt_{12}+\tx_{12}) -\ri \varepsilon \right ]} \S^t_{ij} \Biggr \} \, . \quad 
\label{N2}
\end{eqnarray} 
One easily verifies that the equilibrium limit 
($\beta_i \to \beta$ for all $i$) of (\ref{N2}) satisfies the KMS condition 
\begin{equation}
\langle j_+(t_1,x_1,i) [\varrho_{s+\ri \beta} j_+](t_2,x_2,j) \rangle_{\beta, \mu}^{\rm conn} = 
\langle [\varrho_{s} j_+](t_2,x_2,j) j_+(t_1,x_1,i)  \rangle_{\beta, \mu}^{\rm conn}\, ,
\label{currKMS1}
\end{equation}
where the KMS automorphism $\varrho$ acts on $j_+$ as follows, 
\begin{equation}
[\varrho_s j_+](t,x,i) = j_+(t+s/v,x,i) \, . 
\label{currKMS2}
\end{equation}

The explicit expression (\ref{N2}) contains fundamental physical information about the NESS. 
First of all, since\footnote{The bar indicates complex conjugation.}
\begin{equation} 
\langle j_+(t_1,x_1,i_1) j_+(t_2,x_2,i_2)\rangle_{\beta,\mu}^{\rm conn} \not = \\
\overline {\langle j_+(-t_1,x_1,i_1) j_+(-t_2,x_2,i_2)\rangle}{}_{\beta,\mu}^{\rm conn} 
\label{cc2} 
\end{equation} 
the NESS breaks down {\it time reversal} invariance, even if the junction interaction 
preserves it, i.e. if $\S=\S^t$ \cite{Bellazzini:2009nk}. Nevertheless, 
{\it time translation} invariance is preserved, which allows 
one to use the conventional definition \cite{bb-00} of {\it noise power} 
\begin{equation}
P_{ij}(\beta; x_1, x_2 ; \omega) \equiv \int_{-\infty}^\infty \rd t\,  \e^{i \omega t} \, 
\langle j_x(t,x_1,i) j_x(0,x_2,j) \rangle_{\beta, \mu}^{\rm conn}\, . 
\label{N3}
\end{equation} 
Eq. (\ref{N3}) defines a complex matrix whose entries can be 
expressed \cite{PBM} in terms of the hypergeometric function ${}_2F_1$, namely 
\begin{eqnarray}
P_{ij}(\beta; x_1, x_2 ; \omega) = \left (\frac{v}{2 v_F \zeta_+}\right )^2 
\bigl \{
\bigl [F_-(\omega, \beta_i, x_{12}) - F_+(\omega, \beta_i,x_{12})\bigr ] \beta_i^{-1} \delta_{ij} + 
\qquad \qquad \nonumber \\ 
\sum_{l=1}^n\S_{il}\beta_l^{-1}\bigl [F_-(\omega, \beta_l, -x_{12}) - F_+(\omega, \beta_l, -x_{12})\bigr ]\S^t_{lj} -
\qquad \qquad \qquad \qquad \nonumber \\
\S_{ij}\beta_j^{-1}\bigl [F_-(\omega, \beta_j,-\tx_{12}) - F_+(\omega, \beta_j,-\tx_{12})\bigr ] - 
\bigl [F_-(\omega, \beta_i, \tx_{12}) - F_+(\omega, \beta_i,\tx_{12})\bigr ] \beta_i^{-1} \S^t_{ij} 
\bigl \}\, , \nonumber \\
\label{A1}
\end{eqnarray}
with 
\begin{equation}
F_\pm (\omega, \beta, x) = \frac{\e^{\pm 2\pi x/\beta }}{\ri \omega \beta \pm 2\pi v}\;  
{}_2F_1 \left (2, 1\pm \frac{\ri \omega \beta}{2 \pi v}, 2 \pm  
\frac{\ri \omega \beta}{2 \pi v}, \e^{\pm 2\pi x/\beta }\right )\, . 
\label{A2}
\end{equation}
{}From (\ref{A1}, \ref{A2}) one can deduce 
the zero-frequency limit ({\it zero-frequency noise power}) 
\begin{equation} 
P_{ij}(\beta) \equiv \lim_{\omega \to 0^+} P_{ij}(\beta; x_1, x_2 ; \omega) \, . 
\label{N4}
\end{equation}
Using 
\begin{equation}
\lim_{\omega \to 0^+} \left [F_- (\omega; \beta, x) - 
F_+ (\omega; \beta, x)\right ] = \frac{1}{2\pi v}\, , 
\label{N5}
\end{equation}
one gets  
\begin{equation}
P_{ij}(\beta) = 
\frac{G(g_-,\k)}{v_F}
\left (\beta_i^{-1}\delta_{ij} - \S_{ij}\beta_j^{-1} - \beta_i^{-1} \S^t_{ij} + 
\sum_{l=1}^n \S_{il}\beta_l^{-1} \S^t_{lj} \right )\, , 
\label{N6}
\end{equation}
where $G$, defined by (\ref{ct3}), captures 
the dependence (see Fig.\ref{fig7}) of the noise on the statistical parameter $\k$. 
As expected, $P_{ij}(\beta)$ turns out to be a $x_{1,2}$-independent real symmetric matrix. If the electric 
charge is conserved ($\S \in O_{\bf v}$), the noise power (\ref{N6}) satisfies in addition the Kirchhoff rule 
\begin{equation} 
\sum_{i=1}^n P_{ij}(\beta ) = \sum_{j=1}^n P_{ij}(\beta ) = 0\, .  
\label{N7}
\end{equation}
The expression (\ref{N6}) admits the typical Johnson-Nyquist $\beta^{-1}$ behavior and 
shows the non-trivial interplay between the different temperatures and the scattering matrix. For example, 
in the two-terminal case with $\S=\S^+$ one finds 
\begin{equation}
P^+ = \frac{G}{v_Fk_B}\left(\begin{array}{cc}T_1(1-\cos \theta)^2 +T_2 \sin^2 \theta  & 
(T_1-T_2)(1-\cos \theta)\sin \theta \\ 
(T_1-T_2)(1-\cos \theta)\sin \theta & 
T_2(1-\cos \theta)^2 + T_1\sin^2 \theta \\ \end{array} \right)\, , 
\label{N8}
\end{equation}
where $T=(k_B\beta)^{-1}$ is the absolute temperature 
and $k_B$ is the Boltzmann constant. The eigenvalues of $P^+$ 
\begin{equation}
p^+_i= \frac{2G}{v_Fk_B}(1-\cos \theta)T_i\, \geq 0\, , \qquad 
i=1,2 
\label{N9}
\end{equation}
are nonnegative in agreement with the positivity of the two-point function (\ref{N1}). 
Analogous result holds for $P^-$ corresponding to $\S^-$.

\section{Outlook and conclusions}

In this paper we constructed and investigated an exactly solvable model of a non-equilibrium Luttinger junction. 
The basic points of our approach are: 
\begin{description} 
\item[(i)] a scale invariant point-like interaction, which is described by a scattering matrix 
$\S$ and drives the system away from equilibrium;

\item[(ii)] a representation generated by a NESS $\Omega_{\beta, \mu}$, 
which encodes the point-like interaction in the chiral fields $\phz$; 

\item[(iii)]  an exact operator solution of the TL model (in terms of $\phz$) on a star graph 
with the current splitting boundary condition in the vertex; 

\item[(iv)] an extension of the conventional fermion Luttinger liquid to anyon statistics.  
\end{description}
Combining these ingredients, we derived the basic correlation functions in the 
state $\Omega_{\beta, \mu}$. The essential characteristic features of these functions are: 
\begin{description} 
\item[(a)] the non-equilibrium two-point anyon correlations are products of $\S$-dependent 
powers of equilibrium correlations at the temperatures and chemical potentials of the heat 
baths, connected to the leads;

\item[(b)] accordingly, the corresponding momentum space distribution is the convolution 
of equilibrium anyon distributions at different temperatures and chemical potentials;  

\item[(c)] the Fourier transform of the leading terms in the large distance expansion of the anyon 
correlations gives Cauchy-Lorentz distributions, which after convolution reproduce 
themselves with appropriate width and median; 

\item[(d)] in the critical limit one has a $c=1$ conformal field theory with 
$\S$-dependent anomalous dimensions, which are explicitly derived; 

\item[(e)] the expected breakdown of time reversal invariance is manifest in the current-current 
correlator. 
\end{description}

We investigated in detail the energy and charge transport in the junction for all values 
of the statistical parameter. The energy is conserved for $\S \in O(n)$, 
which covers both possibilities of a junction without and with 
electric charge dissipation. In the latter case we determined the exact expression for the 
charge flow leaving or entering the junction. The connected current-current correlation 
is a linear combination of hypergeometric functions. The associated zero-frequency 
noise power depends linearly on the temperatures. 

Our investigation above has been focused essentially on the critical properties of anyon 
Luttinger liquids away from equilibrium. It will be interesting to study the noncritical 
aspects as well. The generalization of the results of this paper beyond the 
Luttinger liquid paradigm, when the nonlinearity of the dispersion relation becomes 
essential, is also a challenging open problem.  
\bigskip

\leftline{\bf Acknowledgments:}
\medskip 

We thank B. Dou\c{c}ot and I. Safi for an inspiring discussion, which stimulated our interest 
in non-equilibrium Luttinger liquids. M.M. would like also to thank the 
Laboratoire de Physique Th\'eorique d'Annecy-le-Vieux for the kind hospitality during the 
preparation of the manuscript.

\appendix

\section{Large space separation asymptotics}

The behavior of the correlators (\ref{panc1},\ref{panc2}) at large large space 
separation $|x_{12}| \gg \beta_i$ is encoded in   
\begin{eqnarray}
C_{11}(x,i;\beta,\mu) &=& \left (\frac{1}{2\ri}\right )^\k \left (\frac{2\pi}{\beta_i}\right )^\k 
L\left (x;\frac{\pi \k}{\beta_i},-\mu_i\right ) + \cdots \, ,  
\label{ld1} \\ 
C_{22}(x,i;\beta,\mu) &=&  \left (\frac{1}{2\ri}\right )^\k \prod_{k=1}^n \left (\frac{2\pi}{\beta_k}\right )^\k 
L\left (x;\frac{\pi \k\S^2_{ik}}{\beta_k},-\S_{ik}\mu_k\right ) +\cdots \, ,  
\label{ld2}
\end{eqnarray} 
where $x\gg \beta_i >0$, the dots stand for sub-leading contributions and 
\begin{equation}
L(x;\gamma,\mu) \equiv \frac{1}{2\pi} \e^{-\ri \mu x - \gamma x}\, , \quad x>0\, . 
\label{ld3}
\end{equation}
The Fourier transform 
\begin{equation} 
\hL (p;\gamma,\mu) \equiv \int_{-\infty}^\infty \rd x\, \e^{\ri px} L(x;\gamma,\mu) = 
\frac{1}{\pi} \frac{\gamma}{\gamma^2 + (p-\mu)^2} 
\label{ld4}
\end{equation}
is the familiar Cauchy-Lorentz distribution\footnote{Known also as non-relativistic Breit-Wigner distribution.}, where 
$\gamma$ is the half width at half maximum and $\mu$ is the statistical median. Using that the 
class of Cauchy-Lorentz distributions is closed under convolution, one finds 
\begin{eqnarray}
\hC_{11}(p,i;\beta,\mu) &\equiv&  \int_{-\infty}^\infty \rd x\, \e^{\ri px} C_{11}(x,i;\beta,\mu) 
\sim \hL\left (p;\frac{\pi \k}{\beta_i},-\mu_i\right ) + \cdots \, ,  
\label{ld5} \\ 
\hC_{22}(p,i;\beta,\mu) &\equiv&  \int_{-\infty}^\infty \rd x\, \e^{\ri px} C_{22}(x,i;\beta,\mu) 
\sim \hL\left (p;\pi \k\sum_{j=1}^n\frac{\S^2_{ij}}{\beta_j},-\sum_{j=1}^n\S_{ij}\mu_j\right ) +\cdots \, . 
\nonumber \\
\label{ld6}
\end{eqnarray} 

Summarizing, the Fourier transform of the leading term in the long distance expansion 
of both (\ref{panc1}) and (\ref{panc2}) is a Cauchy-Lorentz distribution. Notice that the width and the median 
of (\ref{ld6}) depend on the temperatures and chemical potentials of all heat baths, as well as on $\S$.

\end{document}